\documentclass[a4paper,fleqn,usenatbib]{mnras}

\usepackage[T1]{fontenc}
\usepackage{ae,aecompl}

\usepackage{mathptmx}
\usepackage{graphicx}	% Including figure files
\usepackage{amsmath}	% Advanced maths commands
\usepackage{amssymb}	% Extra maths symbols
\usepackage{lipsum}
\usepackage{times}

\title[CEEs with low-mass giants: the energy budget]
{Common envelope events with low-mass giants: understanding the energy budget}
\author[J.L.A. Nandez and N. Ivanova]{J.L.A. Nandez$^1$\thanks{E-mail:jnandez@sharcnet.ca (JLAN)
}  and N. Ivanova$^2$\\
$^1$SHARCNET, The University of Western Ontario, London, ON, N6A 5B7, Canada\\
$^2$Department of Physics, University of Alberta, Edmonton, AB, T6G 2E7, Canada}

\begin{document}

\date{Draft date: \today}

\pagerange{\pageref{firstpage}--\pageref{lastpage}} %\pubyear{2002}

\maketitle

\label{firstpage}

\begin{abstract}
Common envelope  events are  important interactions between  two binary stars
that lead to the formation of close binary systems.  We present here
a  systematic three-dimensional  study in  which we  model  common
envelope events  with low-mass giant donors.   The results  allow us to
revise the  energy formalism that is  usually used to determine common
envelope event outcomes.  We show that the energy budget for
this type of system should include the recombination energy, and that it 
also  must  take into  account  that  a  significant fraction  of  the
released orbital energy is taken away by the ejecta.  We provide three
ways in which  our results can be used by  binary population synthesis
studies:  a  relation that  links  the  observed post-common  envelope
binary with the initial  binary  parameters,  a fitting  formula  for the 
$\alpha_{\rm  ce}\lambda$ parameter  of  the standard energy  formalism, and a 
revised   energy  formalism   that   takes  into   account  both   the
recombination energy and the energy that is taken away by the ejecta.
\end{abstract}

\begin{keywords}
hydrodynamic --  binaries: close -- white dwarfs -- subdwarfs
\end{keywords}

\section{Introduction}
\label{sec:intro}

It  is  believed that the common  envelope  event  (CEE) is  the  most
important phase in the evolution of a wide range of different types of
close binary  systems. It  most likely  plays a  crucial role  in the
formation of  X-ray binaries,  Type Ia supernovae  progenitors, double
degenerate     stars,     and     more    \citep[for     a     review,
  see][]{2013A&ARv..21...59I}. CEE  is a short-lived  physical process
when two stars orbit inside  a single, shared envelope.  The outcome of
a CEE is either a new binary with  a reduced orbit, or a merger of the
two stars.

One of the standard  ways to predict an outcome of a  CEE is by using the
energy  formalism.  This  method equates  the binding  energy of  the
envelope of  the donor with  the orbital  energy before and  after the
event                          \citep[see                          for
  example][]{1976IAUS...73...35V,1984ApJ...277..355W}:

\begin{equation}
E_{\rm bind}=\Delta E_{\rm orb} \ ,
\end{equation}
where $E_{\rm  bind}$ is  the binding  energy of  the envelope  of the
donor,  and  $\Delta E_{\rm  orb}$  is  the change in the  orbital  energy.
Recognizing that not  all the available orbital energy can  be used to
eject the envelope of  the donor, \citet{1988ApJ...329..764L} proposed
a common-envelope efficiency parameter,  $\alpha_{\rm CE}$, defined as
the fraction of released orbital energy that has been effectively used
to eject the envelope of  the donor.  This $\alpha_{\rm CE}$ parameter
is now widely used in  binary population synthesis studies \citep[see,
  e.g.][]{2002MNRAS.336..449H}.  A better understanding of the
energy budget of  a CEE, better than a simple parameterization, could
help us to better predict the population of close binaries.

The systems where  the parameters of the CEE  can be best constrained  are
double-white-dwarf  (DWD) binaries.   It is  widely believed  that the
last episode  of mass  transfer leading to  DWD formation  was an
unstable MT, a  CEE, where the donor  was a red giant  (RG) star
\citep[e.g.,][]{1981NInfo..49....3T,1984ApJS...54..335I,1984ApJ...277..355W,1988SvAL...14..265T}.  For
RGs,  a well-defined  relation  between their  core  masses and  radii
exists.  From the observations of DWD  systems, we know that one of the
white dwarfs (WDs) is usually younger, and therefore is the remnant of
the  pre-CE RG  donor.  However, the  mass  of the  donor  can not  be
uniquely determined, as long as $\alpha_{\rm CE}$ is uncertain.

In this  paper, we perform three-dimensional  numerical simulations of
CEEs between  low mass RG  stars and WD  companions. This work  is the
extension of our  preliminary study devoted to the  formation of the DWD
WD~1101+364 via  a CEE  \cite{nandez2015}.  Here,  we consider  a wide
parameter space based  on the mass of  the RG donor, the  RG core mass
and  the  companion  mass.   Our  DWD binaries  have a  mass  ratio
$q=M_1/M_2$ between  0.8 and  1.125, where  $M_1$ is  the mass  of the
younger WD (formed during  a CEE), and $M_2$ is mass  of the older WD.
The main goal of this paper is  to understand the energy budget at the
end of  the CEE.  We  pay particular attention to the  usage of the
recombination energy,  and to  the energy  that is  taken away  by the
ejecta.   Those energies  are not  usually taken  into account  by the
standard  energy  formalism,  and  should  explain  the  deviation  of
$\alpha_{\rm CE}$ from 1.

We  describe  the initial  conditions,  the  parameter space  and  the
numerical methods in  \S~\ref{sec:initial}.  \S~\ref{sec:def} contains
the definitions for the  energies.  In \S~\ref{sec:dwdce}, we give an overview of 
the  final states  of the  simulations  in terms  of the  mass of  the
ejecta, energy evolution during a spiral-in, orbital parameters at the end 
of  a  CEE,   and  discuss  how  the  outcomes  of   the  CEE  can  be
parameterized.    Finally,  \S~\ref{sec:conclusion}   gives  a   brief
discussion  on how  our results  can be  used in  population synthesis
studies, as  well as  comparison with the  observed binaries  that are
known to have a post-CE WD.

\section{Parameter space}
\label{sec:initial}

We study the  progenitor systems that have likely  formed the observed
DWDs, in terms of the observed WD masses and the orbital separations.
We adopt  that immediately before  a CEE, the DWD  progenitor binaries
consisted of a low-mass RG with a  core mass close to the observed new
(second-formed) WD, and  of an older (first-formed) WD.   We consider 24
binaries in the parameter space defined by the RG donor mass, the mass
of the newly formed WD (RG core), and the mass of the old WD.  For the
RG donor mass,  $M_{\rm d,1}$, we take $1.2$, $1.4$,  $1.6$, and $1.8$
$M_\odot$.  Each low-mass RG is  considered at two evolutionary points
on its RG branch, namely when  their degenerate He core masses $M_{\rm c,1}$
are $0.32 M_\odot$  and $0.36 M_\odot$ cores (we expect  that the mass
of the new WD will be very similar  to the He core mass of the RG donors).
For the mass of the old WD, $M_{\rm a,2}$, we take $0.32$, $0.36$, and
$0.40$  $M_\odot$,  for  each   case  of  the  RG  donor.   Table
\ref{tab:init} shows  the summary  of the considered  parameter space,
and the initial conditions for each binary.

To create  the initial RG  donor stars,  we use the {\tt  TWIN/EV} stellar
code   \citep[][   recent    updates   are    described   in
  \citealt{2008A&A...488.1007G}]{1971MNRAS.151..351E,1972MNRAS.156..361E}. 
This  allows us to  obtain a realistic initial  one-dimensional (1D)
stellar  profile for  each RG  donor.  Stars  are evolved  until their
degenerate  He cores  have grown  close to  $0.32 M_\odot$,  and $0.36
M_\odot$.

\begin{table*}
\begin{minipage}{178mm}
 \caption{Complete parameter space with initial conditions.}
 \label{tab:init}
 \begin{center}
 \begin{tabular}{lccccccccrcccc}
  \hline
  Model & $M_{\rm d,1}$ & $M_{\rm c,1}$&$M_{\rm a,2}$ &$R_{\rm rlof}$ & $a_{\rm orb,ini}$ & $P_{\rm orb,ini}$ &$E_{\rm pot,ini}$ & $E_{\rm int,ini}$ & $E_{\rm bind,ini}$&$E_{\rm rec,ini}$&$E_{\rm orb,ini}$&$E_{\rm tot,ini}$&$\lambda$\\
  \hline
%interaction between 0.32,0.36, 0.40 and ~0.32 DWD 
  1.2G0.32C0.32D&1.195 & 0.318& 0.32&29.484&59.47 & 43.11 &-24.542&12.214&-12.328&2.725&-1.218&-10.825&1.093\\
  1.2G0.32C0.36D&1.195 & 0.318& 0.36&29.484&60.74 & 44.00 &-24.542 &12.214 &-12.328&2.725 &-1.345&-10.945&1.093\\
  1.2G0.32C0.40D&1.195 & 0.318& 0.40&29.484&61.93 & 44.65 &-24.542 &12.214 &-12.328&2.725&-1.462&-11.066&1.093\\
  
  1.4G0.32C0.32D&1.397 & 0.319& 0.32 &27.735 & 54.47 & 35.52 &-33.772 &16.825 &-16.947&3.369& -1.556 &-15.134&1.217\\
  1.4G0.32C0.36D&1.397 & 0.319& 0.36 &27.735 & 55.59 & 36.24 &-33.772 &16.825 &-16.947&3.369 &-1.715&-15.293&1.217\\
  1.4G0.32C0.40D&1.397 & 0.319& 0.40 &27.735 & 56.64 & 36.82 &-33.772 &16.825 &-16.947&3.369&-1.870&-15.456&1.217 \\
  
  1.6G0.32C0.32D&1.598 & 0.323& 0.32 &25.805 & 49.57 & 29.21 &-45.768 &22.931 &-22.837&4.009& -1.955 &-20.783&1.312\\
  1.6G0.32C0.36D&1.598 & 0.323& 0.36 &25.805 & 50.54 & 29.76 &-45.768 &22.931 &-22.837&4.009& -2.157&-20.985&1.312\\
  1.6G0.32C0.36D-S&1.598 & 0.323& 0.36 &31.250 & 48.61 & 27.97 &-44.769 &22.412 &-22.357&3.997& -2.241&-20.602&1.106\\ 
  1.6G0.32C0.40D&1.598 & 0.323& 0.40 &25.805 & 51.48 & 30.29 &-45.768 &22.931 &-22.837&4.009& -2.353 & -21.181&1.312\\
  
  1.8G0.32C0.32D&1.799 & 0.318& 0.32 &16.336 & 30.77 & 13.59 &-88.123 &43.955 &-44.167&4.676&-3.544 & -43.036&1.401\\
  1.8G0.32C0.36D&1.799 & 0.318& 0.36 &16.336 & 31.37 & 13.86 &-88.123 &43.955 &-44.167&4.676&-3.912&-43.404&1.401\\
  1.8G0.32C0.40D&1.799 & 0.318& 0.40 &16.336 & 31.93 & 14.10 &-88.123 &43.955 &-44.167&4.676&-4.271 &-43.762&1.401 \\
  
%interaction between 0.32,0.36, 0.40 and ~0.36 DWD  
  1.2G0.36C0.32D&1.177 & 0.362& 0.32 &60.088 & 121.63 &127.06&-13.403 &6.649 &-6.754&2.479&-0.587 &-4.861&0.896\\
  1.2G0.36C0.36D&1.177 & 0.362& 0.36 &60.088 & 124.24 &129.46&-13.403 &6.649 &-6.754&2.479&-0.646 &-4.921&0.896\\
  1.2G0.36C0.40D&1.177 & 0.362& 0.40 &60.088 & 126.68 &131.59&-13.403 &6.649 &-6.754&2.479&-0.704 & -4.979&0.896\\
  
  1.4G0.36C0.32D&1.383 & 0.364& 0.32 &56.700 & 111.58 &104.66&-18.200 &9.103 &-9.097&3.135&-0.752 &-6.713&1.037\\
  1.4G0.36C0.36D&1.383 & 0.364& 0.36 &56.700 & 113.88 &106.68&-18.200 &9.103 &-9.097&3.135&-0.829 &-6.790&1.037\\
  1.4G0.36C0.40D&1.383 & 0.364& 0.40 &56.700 & 116.04 &108.49&-18.200 &9.103 &-9.097&3.135&-0.906 &-6.865&1.037\\
  
  1.6G0.36C0.32D&1.592 & 0.363& 0.32 &50.061 &  96.21 & 79.10&-25.439 &12.700 &-12.739&3.830&-1.003 &-9.914&1.163\\
  1.6G0.36C0.36D&1.592 & 0.363& 0.36 &50.061 &  98.13 & 80.64&-25.439 &12.700 &-12.739&3.830&-1.107 & -10.017&1.163\\
  1.6G0.36C0.40D&1.592 & 0.363& 0.40 &50.061 &  99.93 & 82.03&-25.439 &12.700 &-12.739&3.830&-1.208 & -10.118&1.163\\
  
  1.8G0.36C0.32D&1.796 & 0.360& 0.32 &41.147 &  77.51 & 54.32&-37.269 &18.684 &-18.585&4.521&-1.405 &-15.469&1.279\\
  1.8G0.36C0.36D&1.796 & 0.360& 0.36 &41.147 &  79.01 & 55.38&-37.269 &18.684 &-18.585&4.521&-1.551 &-15.614&1.279\\
  1.8G0.36C0.40D&1.796 & 0.360& 0.40 &41.147 &  80.42 & 56.35&-37.269 &18.684 &-18.585&4.521&-1.693 &-15.756&1.279\\
  \hline
 \end{tabular}
 \end{center}
 \medskip
The models  names are composed  as following: two  digits representing
the RG mass are followed by ``G'', three digits representing the RG core mass followed by ``C'', three digits representing the old WD mass followed by ``D''; ``S'' stands for  synchronized case, otherwise the simulation is non-synchronized case.  {$M_{\rm d,1}$, and $M_{\rm c,1}$  are the  total, and core mass  of the  RG, whilst $M_{\rm a,2}$ is the mass of the old WD, in
$M_\odot$}. $R_{\rm  rlof}$ is the radius  of the donor Roche lobe, in
$R_\odot$.  $a_{\rm orb,ini}$  is  the  initial orbital  separation  in
$R_\odot$,  $P_{\rm   orb,ini}$  is   the  initial  orbital   period  in
days. $E_{\rm  pot,ini}$, $E_{\rm  int,ini}$, $E_{\rm  bind,ini}$,  $E_{\rm  rec,ini}$, $E_{\rm  orb,ini}$  and  $E_{\rm tot,ini}$  are  the potential energy of the RG, the internal energy of the RG without recombination, the   binding  energy  of  the   RG  envelope  without
recombination  energy,  the  total  recombination  energy  of  the  RG
envelope,   initial  orbital   energy,  and   initial  total   energy, defined as the sum of the binding, recombination, and initial orbital energies, respectively,  in the  units of  $10^{46}$  erg. $\lambda\equiv -GM_{\rm d,1}(M_{\rm d,1}-M_{\rm c,1})/(E_{\rm bind,ini}R_{\rm rlof})$ is a dimensionless star structure parameter \citep{1990ApJ...358..189D}.
 \end{minipage}
\end{table*}

To  model a  CEE between  a  RG and  a  WD, we  use {\tt  STARSMASHER}
\citep{2010MNRAS.402..105G,2011ApJ...737...49L},  a Smoothed  Particle
Hydrodynamics (SPH)  code.  Technical  details on  using this  code to
model CEEs can be found in \citet{0004-637X-786-1-39}. We 
reiterate the point  made in \citet{nandez2015}, that when a 1D star is
transferred  to a 3D   code  via the  {\it  relaxation}   process  in  {\tt
  STARSMASHER}, the core  of the RG, $M_{\rm c,1}$,  must be increased
slightly by about $0.01\,M_\odot$ so that the resulting profile of the
3D star  for pressure, density,  internal energy and  other quantities
would  match that  of a 1D  star. The RG  envelope is  modeled using  $10^5$
particles, and the RG core  is modeled as a point mass, as  is the WD
(note that a point mass only interacts gravitationally with normal SPH
particles). The envelope mass in our three-dimensional star is $M_{\rm
  env,1}=M_{\rm d,1}-M_{\rm c,1}$.  We found  that for most RGs
with cores  close to $0.4 M_\odot$,  the profiles could not  be matched
well with  1D stars after the  relaxation, and hence RGs with core masses $>0.4
M_\odot$ were excluded from the considered parameter space.  Improving
the  match   of  the  profiles   in  those  more  evolved   donors  is
computationally unfeasible  now, as this  requires such a change  in the
number of  SPH particles and in  their smoothing length, that  the GPU
time would  be increased  by at  least 64  times\footnote{The average
time that is spent on obtaining one model presented in this paper is
  about  1  GPU core  year.   The  global  GPU resource  available  at
  Compute/Calcul Canada  for all  scientists in Canada, in 
  several GPU-equipped clusters, is about 200 GPU core years.}.

The photospheric radius of the star  in SPH, $R_{\rm SPH}$, is defined
as $R_{\rm SPH}=R_{\rm out}+2h_{\rm out}$,  where $R_{\rm out}$ is the
position of the outermost particle  and $h_{\rm out}$ is the smoothing
length of  that particle  \citep[for a detailed  discussion on  how to
  define  the   photospheric  radius  of  a   three-dimensional  star,
  see][]{{0004-637X-786-1-39}}. Defining the  photospheric radius this
way ensures  that all envelope  particles are enclosed  within $R_{\rm
  SPH}$ for the non-synchronized cases.

The   initial  orbital   separation,   $a_{\rm   orb,ini}$,  for   the
non-synchronized  cases is  found  from the  assumption that  $R_{\rm
  SPH}$ is  equal to  the Roche  Lobe (RL)  radius, $R_{\rm  rl}$, and
using     the    approximation     for     the     RL    radius     by
\citet{1983ApJ...268..368E}.   The  initial  orbital  period,  $P_{\rm
  orb,ini}$ is found assuming a Keplerian orbit.

For the synchronized  case, the initial orbital  period and separation
are found at the moment when the outermost particles overfill the donor's
RL  during   the  \textit{scan}  process  \citep[see   \S2.3  of][also
  \citealt{nandez2015}]{2011ApJ...737...49L}. During the scan process,
the envelope's  angular momentum is steadily boosted. This leads
to  the expansion  of  the radius  of  the donor  as  compared to  the
non-rotating  case.  As   a  result,  the  orbital   separation  in  a
synchronized and  a non-synchronized  case may not  match, however,
the  difference in  these quantities  is not  large. The  photospheric
radius is $R_{\rm SPH}=R_{\rm out}$ for the synchronized case.

We use the tabulated  equation of state  (TEOS) incorporated  from {\tt
  MESA} \citep[see \S4.2 of][]{2011ApJS..192....3P} and implemented as
described  in \citep{nandez2015}.   This  TEOS includes  recombination
energy for H, He, C, N, O, Ne, and Mg. The dominant contribution to the recombination energy comes from H, which account for about 59\% of the total energy, followed by He with about 38\%, and 3\% for the rest of the elements, in all our simulations.

\section{Definitions}
\label{sec:def}

In  this  Section  we  declare  definitions  for  the  most  important
quantities.  Definitions   are    adopted   from
\citep{nandez2015}, unless stated otherwise.

\textbf{Energy formalism. } The  energy formalism compares the donor's
envelope binding energy $E_{\rm bind}$  with the orbital energy before
the  CEE, $E_{\rm  orb,ini}$,  and after  the  CEE, $E_{\rm  orb,fin}$
\citep{1984ApJ...277..355W,1988ApJ...329..764L}:
\begin{equation}
\label{eq:standardCE}
E_{\rm  bind}=\alpha_{\rm  bind}   (E_{\rm  orb,fin}-E_{\rm  orb,ini})
\equiv \alpha_{\rm bind} \Delta E_{\rm orb} \ .
\end{equation}
Here  $\alpha_{\rm  bind}$  is  the fraction  of  the  orbital  energy
effectively used to expel the  CE.  This
parameter is equivalent to the commonly used $\alpha_{\rm CE}$, 
and is usually assumed to be $0\le\alpha_{\rm bind}\le 1$.

The potential energy of the donor's envelope in SPH is
\begin{equation}
 E_{\rm pot,ini}=\frac{1}{2}\sum_i m_i \phi_i,
\end{equation}
where $m_i$, and $\phi_i$ are the mass, and specific gravitational energy, respectively, for each SPH particle $i$ in the initial RG profile, including the core. Note that this quantity is computed before the star is placed in the binary configuration. In our SPH method, $\phi_i$ is calculated as in \citet{1989ApJS...70..419H}. 

The internal energy of the donor's envelope in SPH is
\begin{equation}
 E_{\rm int,ini}=\sum_i m_i\left(\frac{3}{2}\frac{kT_i}{\mu_im_{\rm
 H}}+\frac{aT_i^4}{\rho_i}\right),
\end{equation}
where $T_i$,  $\rho_i$ and  $\mu_i$  are the temperature, density, and mean
molecular mass,  respectively, for each  particle $i$ in the initial RG profile. The constants $k$, $a$,  and $m_{\rm H}$ are the Boltzmann constant, radiation constant, and hydrogen atom mass. 

The binding energy of the RG, without the recombination energy, is 

\begin{equation}
E_{\rm bind} = E_{\rm pot,ini}+E_{\rm int,ini}.
\end{equation}

\noindent This binding energy was historically parameterized using the 
parameter $\lambda$ \citep{1990ApJ...358..189D,2013A&ARv..21...59I},

\begin{equation}
\label{eq:lambda}
E_{\rm bind} = - \frac{GM_{\rm d,1}(M_{\rm d,1}-M_{\rm c,1})}{\lambda R}
\end{equation}
\noindent This equation, combined with the energy formalism equation \ref{eq:standardCE}, provides 
the most used equation to find CEE outcomes in binary population synthesis studies, 
where $\alpha_{\rm bind} \lambda$ are used together as one single parameter:

\begin{equation}
\label{eq:alphalambda}
\Delta E_{\rm orb}  = - \frac{GM_{\rm d,1}(M_{\rm d,1}-M_{\rm c,1})}{\alpha_{\rm bind}  \lambda R}
\end{equation}

The orbital energy of the binary system in SPH takes the following form:

\begin{equation}
\label{eq:orbenergy}
 E_{\rm orb}= \frac{1}{2}\mu |V_{12}|^2
   +\frac{1}{2}\sum_i   m_i\phi_i-\frac{1}{2}\sum_j   m_j\phi_j^{\rm
   RL_1}-\frac{1}{2}\sum_k m_k\phi_k^{\rm RL_2},
\end{equation}

\noindent where $\mu=M_1M_2/(M_1+M_2)$ is the reduced mass, and $\vec{V}_{12}=\vec{V}_1-\vec{V}_2$ is the relative velocity of the two stars. The first, second, third and fourth terms give the orbital kinetic energy, the total gravitational energy  of the
binary (with the sum being over  all particles $i$ in  the binary), the self-gravitational energy of the donor (the sum
being over all particles $j$  in star 1), and of the WD (the sum being over all particles $k$
in star 2, initially just the one particle representing the WD), respectively. 

\textbf{Recombination energy. } The recombination energy 
is included in the total value of the specific internal energy provided by TEOS,
and can be found  as
\begin{equation}
 E_{\rm rec,ini}  = \sum_i m_i\left(u_i-\frac{3}{2}\frac{kT_i}{\mu_im_{\rm
 H}}-\frac{aT_i^4}{\rho_i}\right)\equiv \alpha_{\rm  rec}\Delta E_{\rm
 orb}, \label{eq:recene}
\end{equation}

\noindent where  $u_i$ is  the SPH specific  internal energy  for each
particle. $\alpha_{\rm  rec}$ is  the ratio between  the recombination
energy and the released orbital energy. Since the recombination energy
acts  as an  additional  (to  the orbital  energy)  source of  energy,
$\alpha_{\rm  rec}<0$. This  energy is  not  part  of the  usually
considered binding energy, as it is not available immediately, and its
release must be triggered  \citep{2015MNRAS.447.2181I}.  The amount of
stored recombination  energy is proportional to the  mass of the
envelope $E_{\rm rec,ini}= \eta (M_{\rm d,1}-M_{\rm c,1})$. In a fully
ionized  gas that  consists of  only  helium (0.3  mass fraction)  and
hydrogen   (0.7   mass  fraction),   $\eta\simeq1.5\times10^{13}\,{\rm
  erg/g}$.  Our gas chemical composition  is a bit different, and also
our TEOS  takes ionization of  heavier elements into account as well.  For our
donors, we find $\eta\simeq 1.6\times 10^{13}\, {\rm erg/g}$.
{The version of {\tt STARSMASHER} we use evolves, for each SPH particle, the specific internal energy  $u_i$ and density $\rho_i$ \citep[see Eqs. A18 and A7 of][] {2010MNRAS.402..105G}. The pressure is then found from the internal energy, density, and the adopted equation of state.}

\textbf{Total energy. } The total initial energy, 
$E_{\rm tot,ini}$, is 
\begin{equation}
 E_{\rm tot,ini}=E_{\rm orb,ini}+E_{\rm bind}+E_{\rm rec}.
 \label{eq:etotin}
\end{equation}
This quantity is conserved during the evolution of all our models.

\textbf{Bound and  unbound material.  } For  each particle,  its total
energy   is  defined   as  $E_{{\rm   tot,}i}\equiv  0.5   m_iv_i^2  +
m_i\phi_i+m_iu_i$, where  the first,  second and  third terms  are the
kinetic, potential, and internal  energies, respectively.  We classify
our particle as in \citet{nandez2015}:

(i)  the \textit{ejecta},  $m_{\rm unb}$  -- the  particles that  have
positive energy,

(ii) the \textit{circumbinary} material,  $m_{\rm cir}$ -- the  matter that is bound
to the binary, but is located outside of both RLs, and

(iii) the \textit{binary} material, $m_{\rm bin}$ --  the particles that are inside
either of the two RLs.

The total energy of the unbound material at infinity is found when 
the unbound mass is in a steady state after the CEE.  It is computed as
\begin{equation}
 E_{\rm tot,unb}^\infty =  \sum_i E_{\rm tot,i}^{\rm unb}\equiv-\alpha_{\rm unb}^\infty\Delta E_{\rm orb}\ .  \label{eq:totunb}
\end{equation}
\noindent Note that $E_{\rm tot,i}^{\rm unb}$ includes the recombination energy of the unbound material. $\alpha_{\rm unb}^{\infty}$ is the ratio of the energy taken  away by the  unbound material to the released
orbital  energy.

{\bf Final energies.} The total  energy at the end  of the simulation is  distributed in the
``binding''  energy of  the  gas bound  to the  binary,
$E_{\rm  bind,fin}$, the  final orbital  energy of  the binary,  $E_{\rm
orb,fin}$, and the  total energy of the unbound  material at infinity,
$E_{\rm tot,unb}^{\infty}$:
\begin{equation}
 E_{\rm tot,fin}=E_{\rm orb,fin}+E_{\rm bind,fin}+E_{\rm tot,unb}^{\infty},
\label{eq:totefinal}
\end{equation}
where $E_{\rm tot,unb}^{\infty}$ is  composed of $E_{\rm kin,unb}^{\infty}$,
$E_{\rm int,unb}^{\infty}$, and $E_{\rm pot,unb}^{\infty}$ -- the kinetic,
internal and potential energies of the unbound material, respectively.

Generally,  $E_{\rm bind,fin}$  has a
fairly small absolute value at the end of  the simulation, and so can be safely
disregarded. In addition, the particles  around the WD may be accreted
during a CEE, and hence their presence there may not have any physical
meaning.     The    extended    energy    formalism,    following    to
\cite{nandez2015}, can then be written as follows:
\begin{equation}
 \alpha_{\rm bind}+\alpha_{\rm rec}+\alpha_{\rm unb}^\infty \approx1,
\label{eq:newform}
\end{equation}

\noindent   If $\alpha_{\rm  rec}=\alpha_{\rm unb}^\infty=0$,  
then Equation~\ref{eq:newform} reduces to  the standard  energy formalism. 

For  additional analysis  of the  energies  at the  end of  a CEE,  we
introduce 3 more quantities :
\begin{itemize}
 \item $\alpha_{\rm pot}\equiv E_{\rm pot,unb}^\infty/E_{\rm pot,ini}$
   -- the ratio of  the potential energy taken away by  the ejecta, to
   the initial potential energy of the RG envelope,
 \item    $\alpha_{\rm   th}\equiv    E_{\rm   int,unb}^\infty/(E_{\rm
   int,ini}+E_{\rm  rec})$  --  the   ratio  of  the  internal  energy
   (including recombination) taken  away by the ejecta, to  the sum of
   the initial internal energy and  the recombination energy of the RG
   envelope,
 \item  $\alpha_{\rm kin}^\infty\equiv  -E_{\rm kin,unb}^\infty/\Delta
   E_{\rm orb}$ --  the ratio of the kinetic energy  taken away by the
   ejecta, to the released orbital energy.
\end{itemize}

We point  out that $\alpha_{\rm  kin}^\infty$ is a part  of $\alpha_{\rm
  unb}^\infty$, however, $\alpha_{\rm pot}$  and $\alpha_{\rm th}$ are
not a part  of $\alpha_{\rm unb}^\infty$  as they describe  fractions of
their corresponding initial energies.

All our  simulations conserved quite  well the total  angular momentum
and the total energy.  We have checked and found that the error in the
energy conservation in  all our simulations is less than  0.1\% of the
initial  total  energy,  while  the  error  in  the  angular  momentum
conservations in all  our non-synchronized cases is  less than 0.001\%
of the  initial total  angular momentum  value, and  the error  in the
angular momentum conservation in the only synchronized case is 0.4\%.

\section{Results}
\label{sec:dwdce}

\subsection{Overview}

{\bf Masses.}   At the end  of each simulation, we  form a binary 
consisting of  $M_1$ and  $M_2$ (see Table~\ref{tab:DWD}).
We note  that  $M_1$ and  $M_2$  in Table  \ref{tab:DWD} 
differ from the  values given for $M_{\rm c,1}$, and  $M_{\rm a,2}$ in
Table \ref{tab:init}, respectively, as a  few SPH gas particles remain
within the  RLs of the  DWD binary.  Ultimately, the  ejected material
$M_{\rm unb}$ is at least 99.4\% of the initial RG envelope, and there
is no circumbinary matter around the newly formed DWD binary.

\begin{table*}
\begin{minipage}{165mm}
 \caption{Energies and masses}
 \label{tab:DWD}
 \begin{center}
 \begin{tabular}{lcccccccccccc}
  \hline
  Model & $M_{\rm unb}$ & $M_{1}$ & $M_2$ & $E_{\rm kin,unb}^{\infty}$ & $E_{\rm int,unb}^{\infty}$ & $E_{\rm pot,unb}^{\infty}$&$E_{\rm tot,unb}^{\infty}$&$E_{\rm orb,fin}$ & $E_{\rm bind,fin}$ & $E_{\rm tot,fin}$&$\Delta E_{\rm orb}$ \\
  \hline
  1.2G0.32C0.32D&0.870&0.324&0.320&4.827&0.757&-0.044&5.539&-15.653&-0.712 &-10.826&-14.308\\
  1.2G0.32C0.36D&0.872&0.323&0.360&4.604&0.629&-0.041&5.192&-15.504&-0.639 &-10.951&-14.159\\
  1.2G0.32C0.40D&0.872&0.323&0.400&7.094&1.182&-0.069&8.206&-18.847&-0.430 &-11.071&-17.385\\
  
  1.4G0.32C0.32D&1.074&0.323&0.320&3.733&1.490&-0.218&5.006&-19.638&-0.510 &-15.142&-18.082\\
  1.4G0.32C0.36D&1.079&0.319&0.360&6.790&0.907&-0.094&7.603&-22.911&-0.005 &-15.313&-21.196\\
  1.4G0.32C0.40D&1.074&0.323&0.400&5.797&1.688&-0.248&7.237&-22.329&-0.364 &-15.456&-20.459\\
  
  1.6G0.32C0.32D&1.271&0.323&0.324&4.212&2.030&-0.475&5.767&-26.153&-0.406 &-20.792&-24.198	\\
  1.6G0.32C0.36D&1.274&0.323&0.362&6.074&1.044&-0.155&6.964&-27.814&-0.145 &-20.995&-25.657\\
  1.6G0.32C0.36D-S&1.274&0.323&0.362&6.205&0.686&-0.093&6.798&-27.292&-0.145 &-20.639&-25.051\\
  1.6G0.32C0.40D&1.274&0.323&0.401&7.115&1.316&-0.205&8.226&-29.273&-0.140 &-21.187&-26.920\\
  
  1.8G0.32C0.32D&1.481&0.318&0.320&8.753&3.532&-1.277&11.008&-53.454&-0.621&-43.067&-49.910\\
  1.8G0.32C0.36D&1.478&0.318&0.362&8.333&1.675&-0.371&9.637 &-52.873&-0.171&-43.407&-48.961\\
  1.8G0.32C0.40D&1.479&0.318&0.402&7.990&2.755&-0.934&9.811 &-53.115&-0.729&-43.768&-48.609\\
  
  %interaction between 0.32,0.36, 0.40 and ~0.36 DWD  
  1.2G0.36C0.32D&0.808&0.370&0.320&2.652&0.693&-0.042&3.303 &-7.641&-0.526 &-4.864&-7.054\\
  1.2G0.36C0.36D&0.808&0.368&0.360&1.896&0.985&-0.089&2.792 &-7.200&-0.514 &-4.922&-6.554\\
  1.2G0.36C0.40D&0.808&0.369&0.400&3.811&0.449&-0.021&4.239 &-8.781&-0.437 &-4.979&-8.077\\
  
  1.4G0.36C0.32D&1.013&0.370&0.320&2.863&0.886&-0.074&3.675&-9.963&-0.428 &-6.716&-9.211\\
  1.4G0.36C0.36D&1.013&0.370&0.360&2.498&1.282&-0.141&3.639&-9.994&-0.437 &-6.792&-9.165\\
  1.4G0.36C0.40D&1.013&0.371&0.400&2.842&1.155&-0.109&3.888&-10.249&-0.508 &-6.869&-9.343\\
  
  1.6G0.36C0.32D&1.229&0.363&0.320&4.111&0.904&-0.067&4.948&-14.851&-0.013 &-9.916&-13.848\\
  1.6G0.36C0.36D&1.222&0.363&0.360&4.009&1.766&-0.238&5.537&-15.512&-0.047 &-10.022&-14.405\\
  1.6G0.36C0.40D&1.224&0.368&0.400&3.773&1.687&-0.226&5.234&-14.993&-0.366&-10.125&-13.785\\
  
  1.8G0.36C0.32D&1.436&0.360&0.320&3.990&2.392&-0.425&5.957&-21.233&-0.195&-15.471&-19.828\\
  1.8G0.36C0.36D&1.436&0.360&0.360&4.407&3.700&-1.371&6.735&-21.727&-0.623&-15.615&-20.176\\
  1.8G0.36C0.40D&1.433&0.360&0.403&4.852&2.258&-0.448&6.661&-22.047&-0.380&-15.766&-20.354\\
  \hline
 \end{tabular}
 \end{center}
 \medskip
 $M_{\rm unb}$, $M_1$,  and $M_{2}$ are the unbound,  stripped RG core
 and  old   WD,  in  $M_\odot$.  $E_{\rm   kin,unb}^{\infty}= \sum_i m_i^{\rm unb} v_i^2/2$, 
 $E_{\rm   int,unb}^{\infty}=\sum_i m_i^{\rm unb} u_i$,   $E_{\rm   pot,unb}^{\infty}=\sum_i m_i^{\rm unb} \phi_i$,   and   $E_{\rm
   tot,unb}^{\infty}$  are  kinetic,  internal,  potential  and  total
 energies,  respectively,  for  the   unbound  material.  $E_{\rm orb,fin}$  is  the  orbital energy  after  the CEE.
 $E_{\rm bind,fin}$ is the total energy  of the particles that remained bound to the binary.
$E_{\rm  tot,fin}$ is  the total  energy of all the particles, and $\Delta E_{\rm orb}$ is the released orbital energy. All energies are in $10^{46}$ erg.
 \end{minipage}
\end{table*}

{\bf Final Time.} We stop our simulations no less than 800 orbits after the end of the plunge-in, and typically we stop the simulations after more than 2000 orbits. The plunge-in is the fastest phase of the spiral-in, during which the instantaneous separation (distance) between the RG core and the WD changes substantially on the timescale comparable to its inferred orbital period. At the moment we stop, the orbital separation is changing by less than  $|\delta a_{\rm orb} / a_{\rm orb}| < 0.002$, where $\delta a_{\rm orb}$  is found over one binary orbital period. Some simulations were calculated for much longer, e.g. the case 1.8G0.32C0.36D is calculated for more than 10 000 orbits after the end of the plunge-in, and the change of the orbital separation over the binary orbital period, at the end of the simulation, is $|\delta a_{\rm orb} / a_{\rm orb}| \approx 0.0002$. The final parameters provided in Table~\ref{tab:DWD} are expected to be time-converged values.

\begin{figure}
 \includegraphics[scale=0.4]{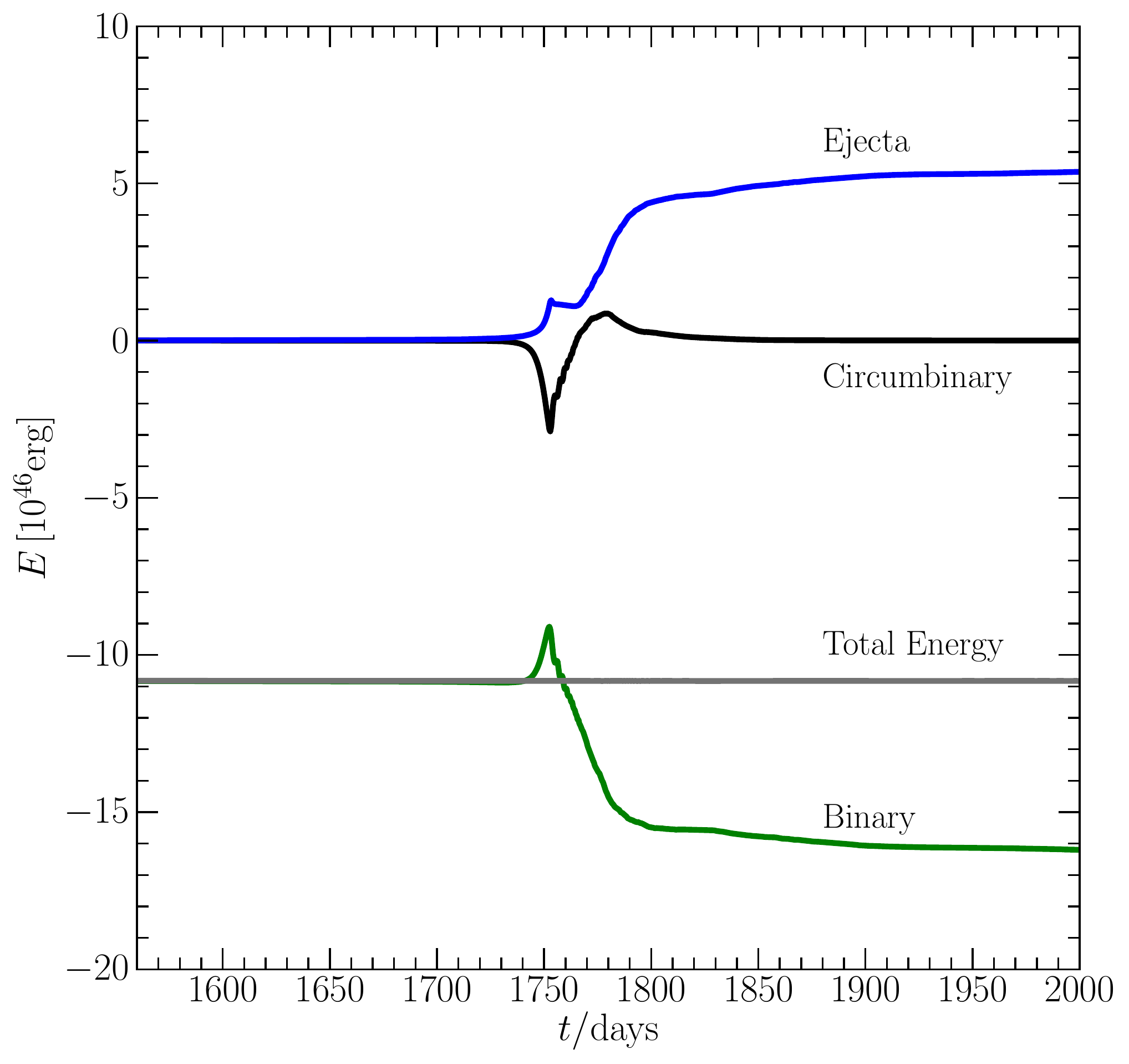}
 \caption{The evolution  of the  total energy  (grey solid  line), the
   orbital energy $E_{\rm orb}$ (green  solid line), the energy in the
   circumbinary matter (black solid line), and the energy in the ejecta (blue
   solid line) for the case 1.2G0.32C0.32D.}
 \label{fig:ene_evol}
\end{figure}

{\bf  Energies.}  Figure  \ref{fig:ene_evol}  shows  how the  energies
change during the spiral-in phase  for the case 1.2G0.32C0.32D.  After
the spiral-in phase is complete and there is no circumbinary
matter left, the circumbinary  total energy vanishes, while  the ejecta energy and
binary total energy (which is the  orbital energy plus the ``binding'' energy
of the remaining  particles) converge to  their final values.
Table \ref{tab:DWD}  provides the final distributions  of energies for
all our simulations.
Some ostensible deviations can be observed in Table~\ref{tab:DWD}. For example,
  the model 1.8G0.32C0.32D has more energetic ejecta than other models. We note that the overall energy budget, and the energy that was extracted from the formed binary, are much higher in this model than in any other model.

\begin{figure}
 \includegraphics[scale=0.4]{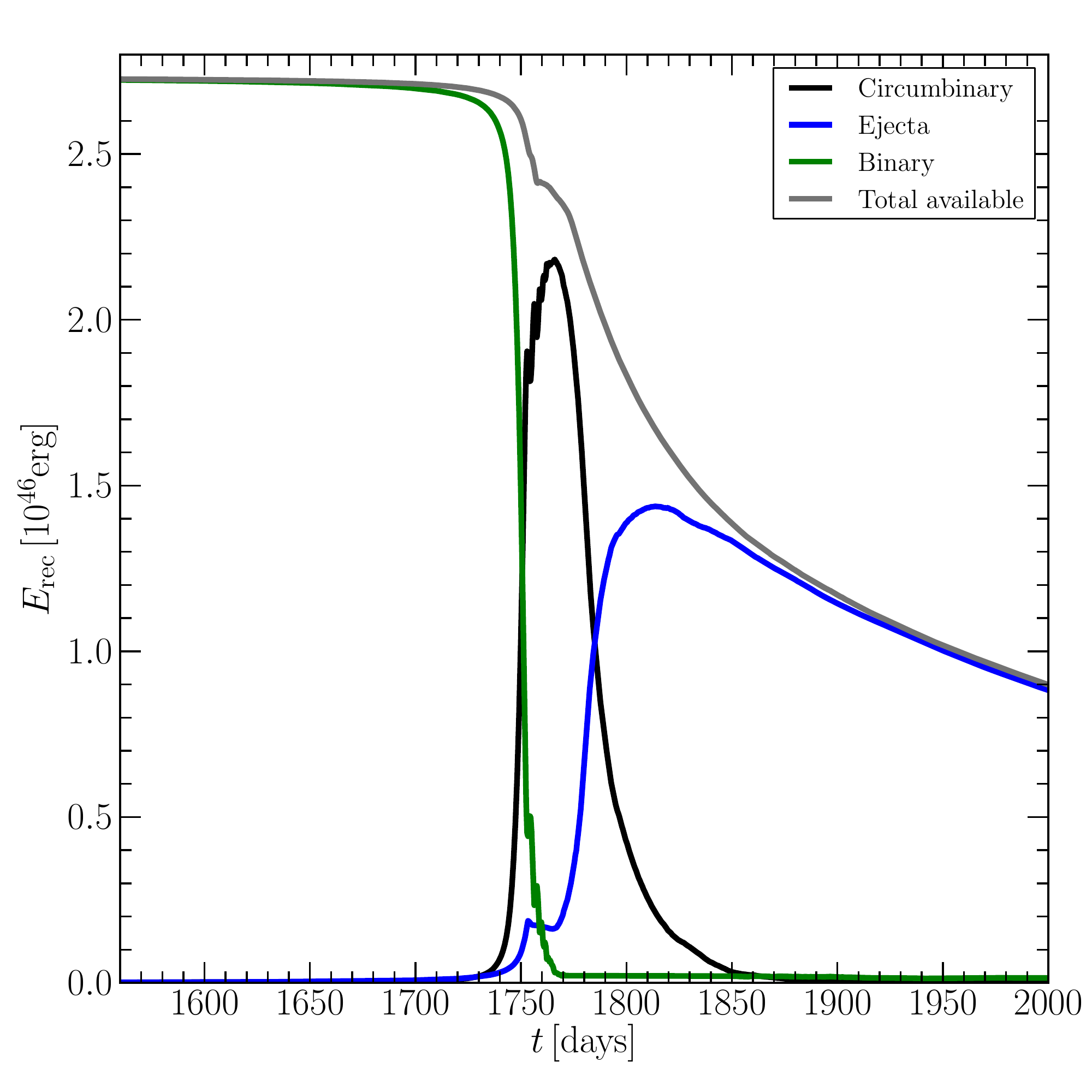}
 \caption{The evolution of the  total recombination energy (grey solid
   line),  the recombination  energy  in SPH  particles  bound to  the
   binary  (green  solid  line),   the  recombination  energy  in  SPH
   particles  in   the  circumbinary material (black  solid   line),  and  the
   recombination energy  in the ejecta  (blue solid line  line), shown
   for the case 1.2G0.32C0.32D.}
 \label{fig:rec}
\end{figure}

\begin{table*}
\begin{minipage}{175mm}
 \caption{Orbital parameters}
 \label{tab:DWDj}
 \begin{center}
 \begin{tabular}{lccccccccccccc}
  \hline
  Model &$r_{\rm p}$ & $r_{\rm a}$ & $a_{\rm orb,fin}$ & $P_{\rm orb,fin}$ & $e$ &$\alpha_{\rm bind}$&$\alpha_{\rm rec}$&$\alpha_{\rm unb}^{\infty}$&$\alpha_{\rm pot}$&$\alpha_{\rm th}$&$\alpha_{\rm kin}^{\infty}$&$\alpha_{\rm bind}\lambda$\\
  \hline
  1.2G0.32C0.32D&1.342&1.415&1.379&0.234&0.026 &0.856&-0.189&0.384&0.003&0.064&0.323&0.936\\
  1.2G0.32C0.36D&1.520&1.532&1.526&0.264&0.004&0.871&-0.192&0.367&0.002&0.042&0.325&0.952\\
  1.2G0.32C0.40D&1.415&1.422&1.419&0.230&0.002 &0.709&-0.157&0.472&0.003&0.079&0.408&0.775\\
  
  1.4G0.32C0.32D&1.093&1.153&1.123&0.172&0.027&0.937&-0.186&0.277&0.006&0.074&0.206&1.140\\
  1.4G0.32C0.36D&1.070&1.089&1.080&0.158&0.009&0.800&-0.159&0.359&0.003&0.045&0.320&0.974\\
  1.4G0.32C0.40D&1.197&1.243&1.220&0.184&0.019&0.828&-0.165&0.354&0.007&0.084&0.283&1.008\\
  
  1.6G0.32C0.32D&0.848&0.882&0.865&0.116&0.020&0.944&-0.166&0.238&0.010&0.075&0.174&1.239\\
  1.6G0.32C0.36D&0.880&0.948&0.914&0.122&0.037&0.893&-0.157&0.273&0.003&0.040&0.237&1.172\\
  1.6G0.32C0.36D-S&0.912&0.947&0.930&0.126&0.019&0.895&-0.160&0.277&0.002&0.026&0.248&1.000\\
  1.6G0.32C0.40D&0.936&0.979&0.958&0.128&0.022&0.848&-0.149&0.306&0.005&0.049&0.264&1.113\\
  
  1.8G0.32C0.32D&0.409&0.448&0.429&0.041&0.046&0.895&-0.095&0.226&0.022&0.091&0.174&1.254\\
  1.8G0.32C0.36D&0.464&0.493&0.479&0.047&0.030&0.902&-0.096&0.197&0.004&0.034&0.170&1.264\\
  1.8G0.32C0.40D&0.517&0.534&0.526&0.052&0.017&0.909&-0.096&0.202&0.018&0.071&0.163&1.274\\
  
  %interaction between 0.32,0.36, 0.40 and ~0.36 DWD  
  1.2G0.36C0.32D&3.174&3.316&3.245&0.816&0.022&0.957&-0.351&0.468&0.003&0.076&0.376&0.857\\
  1.2G0.36C0.36D&3.625&3.834&3.730&0.978&0.028&1.031&-0.378&0.426&0.007&0.108&0.289&0.924\\
  1.2G0.36C0.40D&3.346&3.596&3.471&0.855&0.036&0.836&-0.307&0.525&0.002&0.049&0.472&0.749\\
  
  1.4G0.36C0.32D&2.516&2.559&2.538&0.564&0.009&0.988&-0.340&0.399&0.004&0.072&0.311&1.027\\
  1.4G0.36C0.36D&2.736&2.866&2.801&0.636&0.023&0.993&-0.342&0.397&0.008&0.105&0.273&1.032\\
  1.4G0.36C0.40D&2.911&3.080&2.996&0.685&0.028&0.974&-0.336&0.416&0.006&0.094&0.304&1.012\\
  
  1.6G0.36C0.32D&1.627&1.773&1.700&0.311&0.043&0.920&-0.277&0.357&0.003&0.055&0.297&1.070\\
  1.6G0.36C0.36D&1.777&1.870&1.824&0.336&0.026&0.884&-0.266&0.384&0.009&0.107&0.278&1.028\\
  1.6G0.36C0.40D&2.039&2.143&2.091&0.400&0.025&0.924&-0.278&0.380&0.009&0.102&0.274&1.075\\
  
  1.8G0.36C0.32D&1.129&1.234&1.182&0.181&0.045&0.937&-0.228&0.300&0.011&0.103&0.201&1.200\\
  1.8G0.36C0.36D&1.209&1.333&1.271&0.196&0.049&0.921&-0.224&0.334&0.037&0.160&0.218&1.178\\
  1.8G0.36C0.40D&1.379&1.458&1.419&0.224&0.029&0.913&-0.222&0.327&0.012&0.097&0.238&1.168\\
  \hline
 \end{tabular}
 \end{center}
 \medskip
The closest and farthest orbital separations are $r_{\rm p}$ and $r_{\rm a}$, respectively, while $a_{\rm  orb,fin}$ is  the
semimajor axis (all in $R_{\odot}$). The orbital period $P_{\rm orb,fin}$ is given  in  days, and  $e$  is  the  eccentricity  of  the
 orbit.  The energy fractions $\alpha_{\rm  bind}$,  $\alpha_{\rm rec}$,  and  $\alpha_{\rm
   unb}^\infty$   are defined   in
 Eq.      \ref{eq:standardCE},      Eq.      \ref{eq:recene},      and
 Eq. \ref{eq:totunb}, respectively. $\alpha_{\rm pot}$ is the fraction of potential energy taken by the ejecta with respect to the initial potential energy. $\alpha_{\rm th}$ is the ratio of the thermal energy taken by the ejecta to the initial thermal energy (including recombination energy) and $\alpha_{\rm kin}^{\infty}$ is the kinetic energy scaled with the released orbital energy.     
 \end{minipage}
\end{table*}

\begin{figure}
 \includegraphics[scale=0.4]{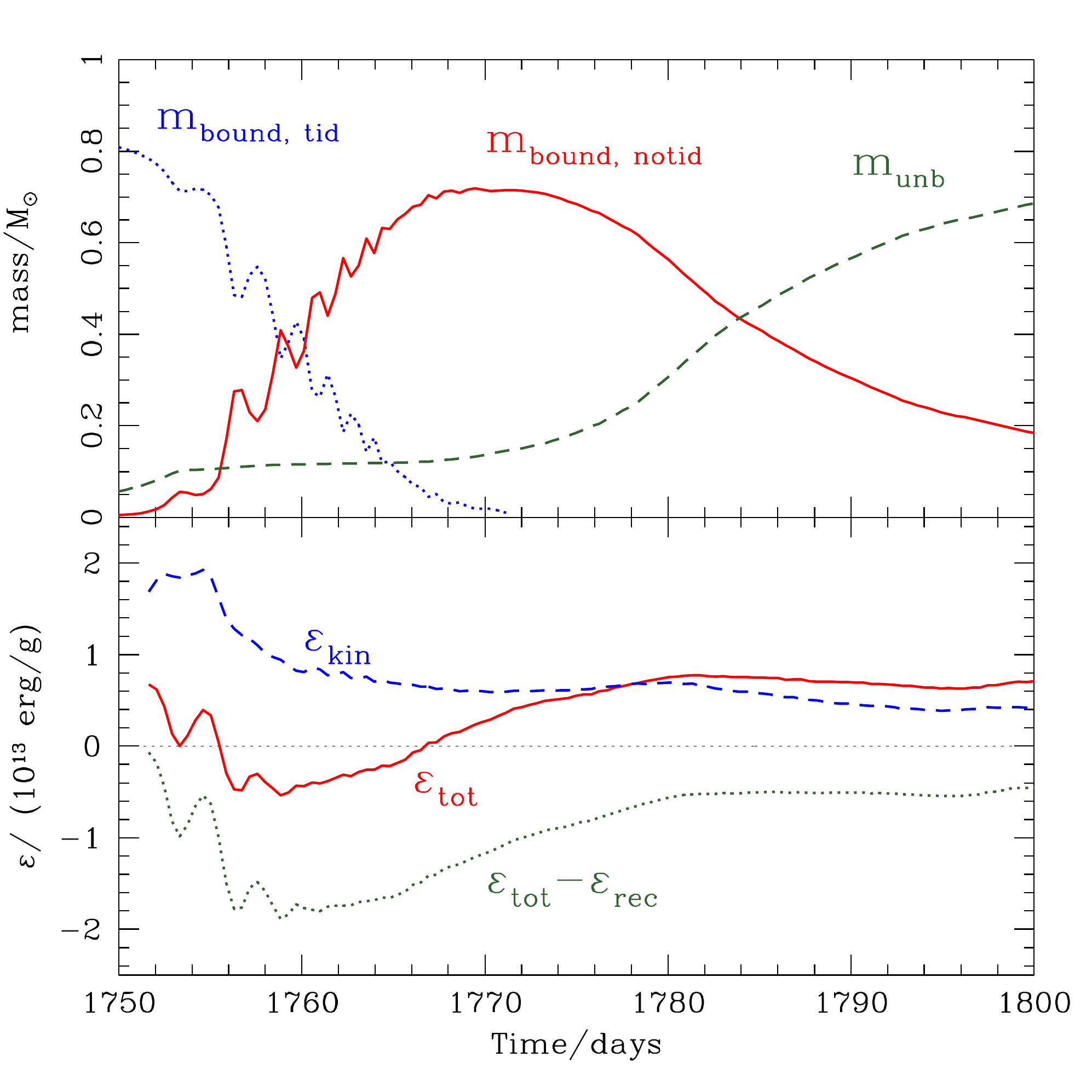}
 \caption{The evolution of the specific energies and the masses for the case 1.2G0.32C0.32D.   
   The bottom plot shows the specific energies per unit mass in the outer envelope (see \S~4.2): kinetic energy $\varepsilon_{\rm kin}$, total energy including recombination $\varepsilon_{\rm tot}$, and  total energy excluding recombination $\varepsilon_{\rm tot}-\varepsilon_{\rm rec}$. See the text in \S4.2 for details on masses.
 }
 \label{fig:rec_inner}
\end{figure}

\subsection{Role of the recombination  energy.   }

\label{ss_rec}

In brief, we  found  that  the  circumbinary
recombination energy has a maximum during the plunge-in phase. At this
moment, almost the entire envelope is  no longer within the Roche lobe
of the binary and becomes circumbinary material
(see an example in Figure \ref{fig:rec}).
The  ejection of  the circumbinary  envelope then  takes place  on its
dynamical  time-scale, which,  e.g.  for  the case  1.2G0.32C0.32D, is
about 100 days.

Let us consider in detail why the dominant energy source that drives the final ejection of the puffed-up envelope
  is the recombination energy but not the binary orbital energy.
  For that, we will trace
  the evolution of {\it specific energies} during the crucial time at about the plunge-in,
  when the puffed-up circumbinary envelope formed initially.

The binary orbital energy can be expected to boost the envelope's ejection by being transferred 
into kinetic energy of the envelope. However, the acceleration of material by the binary's motion, which also can be called 
a dynamical tide, will only affect the mass that is located approximately within 3 binary orbital separations
 of the binary's center of mass \citep[see, e.g.][]{1993A&A...280..174P}.
Note that we do not consider here any secular tidal effects, but only 
an acceleration that is produced during a period of time that is less than
a hundred orbital periods of the binary after the plunge-in,
i.e. comparable to when the envelope is ejected in our simulations.
We can test whether the mass located far away from the binary is, or is not, accelerated by the binary in our case.

First, we separate the bound envelope into two sub-envelopes:
the inner envelope - the envelope's
mass that is within the distance of $3a_{\rm orb}$, $m_{\rm bound,tid}$ -- and
the outer envelope,  $m_{\rm bound,notid}$, or the mass that is beyond $3a_{\rm orb}$.
Here $a_{\rm orb}$ is the current distance between
the RG core and the WD.
The outer envelope, once it is ``decoupled'' from the binary's tidal effect,
is expected to evolve according to its potential, kinetic, internal and recombination energies.
  
Let us consider the case  1.2G0.32C0.32D (see Figure \ref{fig:rec_inner}).
In the evolution shown, the plunge-in takes place from days 1750 to 1770, when
the orbital separation shrinks by a factor of 10,
approaching closely its final value. At day 1770,
most of the initial envelope is either in the outer envelope, or is ejected.
During the plunge-in, the definition of the orbital separation  $a_{\rm orb}$
by the energetic principle can not provide a proper result \citep[for the energy budget and the orbital separation
ambiguity during the plunge-in see][]{ivanova16}, hence 
we can only use a geometrical distance between the RG core and WD.
During the plunge-in, the inferred orbital separation is changing rapidly (in a sense, it
can also be described as having a very high eccentricity).
Since the boundary between the outer and inner envelopes, drawn at $3a_{\rm orb}$, oscillates as well as the orbital separation itself, the defined masses of the envelopes and their energies oscillate during the plunge-in.

In Figure \ref{fig:rec_inner}, we can see that the specific kinetic energy of the
outer envelope is settled by the  end of the plunge-in.
This outer envelope is bound by the conventional definition,
in which the total energy excluding recombination is negative.
At the same time, the outer envelope is effectively decoupled
from the binary and is not receiving further boosts to its kinetic energy.

At the moment of the end of the plunge-in, the outer envelope possesses
 most of the mass that remains bound to the binary.
The outer envelope has obtained some kinetic energy from its previous interaction
with the shrinking binary during the plunge-in.
That non-zero kinetic energy leads to the envelope expansion, on the dynamical timescale of the expanded envelope,
where every SPH particle in the outer envelope can have only a parabolic (bound) trajectory
with respect to the binary, if the recombination energy is not released.

However, once the material expands and cools down 
enough to start recombination, an SPH particle gains enough energy to become unbound -- 
it can be seen from Figure \ref{fig:rec_inner} that at the end of the plunge-in
the stored potential recombination energy is sufficient to unbind the material of the outer envelope.
The outer envelope is now flowing away; 
as more of its material feels a pressure differential between open space above and the remaining envelope below,
it expands further, cools down and becomes unbound after recombination.
This recombination-driven ejection is gradual and non-explosive, vs. the rather explosive,
or dynamical, ejection that takes place  during the plunge-in, as  described in detail by \cite{ivanova16}.
The radius at which the released recombination energy can remove a particle out
of the potential well is  the ``recombination'' radius, and was derived in \citep{ivanova16}.
We can clarify that there is  no recombination energy  stored in  the ejected material  at  the end  of  the  simulations. 

It is important to mention that the recombination takes place at large  
optical depths. Using our 3D models, we estimate that typical optical depths have values of at least 10,
and 1D studies showed that hydrogen recombination can take place at optical depths
above 100 \citep{2015MNRAS.447.2181I}.

\subsection{Post-CE orbital  parameters}   

We find the final orbital  separation in a \textit{geometrical way} as
$a_{\rm orb,fin}=(r_{\rm  a}+r_{\rm p})/2$,  where $r_{\rm p}$  is the
periastron, and $r_{\rm a}$ is the apastron.  We ensure that these two
quantities, $r_{\rm  p}$ and $r_{\rm a}$, are no longer changing with  time at
the moment  when we extract  them from  the simulations.  We  find the
final orbital period of the binary assuming a Keplerian orbit, $P_{\rm
  orb,fin}$.  Another important orbital parameter is the eccentricity,
$e$, which  is found as $e=(r_{\rm  a}-r_{\rm p})/(2a_{\rm orb,fin})$.
The final orbital parameters are provided in Table \ref{tab:DWDj}.

Note that the final  separation  found using  the orbital
energy, $a_{\rm  orb,fin}^{\rm En}=-GM_1M_2/(2E_{\rm  orb,fin})$,  differs  from  the  final  orbital
separation found in the geometric way, $a_{\rm orb,fin}$. This is for two
reasons:

\begin{enumerate}

 \item There is  still mass within the  RLs of both  point masses, as
 can be seen by the non-zero value of $E_{\rm bind, fin}$. The presence of
 these particles, and  their not fully stable orbits  around the point
 masses, makes the energy-based  way to calculate $a_{\rm orb,fin}^{\rm
   En}$  uncertain.  Note  that  these few  particles  make the  stars
 aspherical  and  the  equation  for  $a_{\rm  orb,fin}^{\rm  En}$  is
 formally not valid.

 \item The distance between the two point masses (WD and RG core) is less
 than two  times their smoothing  lengths, which means that  there is 
 some  extra   smoothing  in  the  gravitational   potential  equation
 \citep[see the Appendix of ][]{1989ApJS...70..419H}.
 The smoothing length of the point masses acts as the softening term defined  by
   \citet{1989ApJS...70..419H}.
   For details on the definition and how to determine the smoothing length in the case modeled here,  see \citet{2011ApJ...737...49L}.
As an example, in the model 1.8G0.32C0.32D, the smoothing length for the RG core is  $h_{\rm core}=0.35 R_\odot$, 
and the smoothing length for the WD is $h_{\rm WD}=0.73 R_\odot$.

\end{enumerate}
 
 The  difference in  orbital separations  between the  geometrical way
 $a_{\rm  orb,fin}$  and the  energy  way  $a_{\rm orb,fin}^{\rm  En}$
 varies  from  7.19\%  (1.2G0.32C0.36D) to  18.11\%  (1.8G0.32C0.32W),
 where the separation derived via the geometrical way is always
 smaller than the separation  derived via the energy way.
There is  a very small discrepancy for  the initial orbital separations  using the two
methods, $<0.24$\%.

The two values for the orbital separation
  would be closer to each other if the potential in the SPH code were calculated without
  a softening term. Note however that due to the first reason above,
  which is the presence of SPH particles inside the RLs,
  the two terms will never be completely the same.
  The discrepancy between the two values due to the softening term is expected
  to decrease if the smoothing length is decreased, and that can be done if the number of the particles is increased,
  although it is not intuitive to state whether the separation found by the geometrical way would increase or decrease.
  Only one test was made for the CEE study of the formation of the specific binary,
  a simulation with 200k particles  resulted  in a 7 per cent smaller final orbital
  separation than the same case  modelled with 100k particles \citep{nandez2015}.
  To clarify, the smoothing length was smaller in the case of 200k by 20 per cent compared to the case of 100k,
  but the relative difference between the final separations derived in the two ways was smaller.
  The models presented in this study might be affected similarly, but it is likely that the relative change in the final
  results will be small even if the resolution will be doubled.
   
Even though the smoothing lengths of the point masses are  
partially responsible for the discrepancy between the orbital separations found by the two methods,
the smoothing length values cannot explain the unbinding of the puffed-up envelope.
For instance, let us consider the model mentioned above with
the maximum discrepancy of 18\%, 1.8G0.32C0.32W. The smoothing lengths for the RG core
and WD are $h_{\rm core}=0.35 R_\odot$, and $h_{\rm WD}=0.73 R_\odot$.
The softening in this case starts to work when the distance between
the RG core and the WD is $2.16R_\odot$. At that moment, most of the mass
is located at an average distance of 14~$R_\odot$, except for a few strongly bound particles which remain
bound within about 3~$R_\odot$ from the center of mass. The final separation therefore
can be dependent on the mass resolution of the particles that were initially strongly bound and were
in the close neighborhood of the RG core (where the smoothing length become important).
However, as was discussed previously in \S~\ref{ss_rec}, the binary is decoupled with the puffed-up outer envelope, which is
too far from the binary to be able to effectively transfer away  its orbital energy,
and the envelope ejection only depends on the stored kinetic and recombination energy, and does not affect
the final parameters of the binary.

Table \ref{tab:DWDj} shows that the bigger the initial mass of the RG,
the tighter  is the final orbit,  for each fixed companion  mass.  For
each  initial  RG mass  and  different  companion mass,  usually,  the
smaller the mass of the companion, the tighter the final orbital
separation.  However, there are two exceptions:

\begin{itemize}
 \item In  the case of the 1.2$\, M_\odot$ RG with a 0.32$\,M_\odot$ core,
   the  largest  final   orbital  period  is  for   the 
   0.36$\,M_\odot$ WD companion,  instead of the 0.40$\,M_\odot$ WD.  This  could be
   because  the  1.2G0.32C0.36D  case  deposited the {\it  least}  of  the
   kinetic energy  in the ejecta, as  compared to the other  two cases
   during  the  spiral-in   phase  (see  Table  \ref{tab:DWD}).

 \item In  the case of the 1.4$\, M_\odot$ RG with a 0.32$\,M_\odot$ core,
   the tighter final  orbit is for the  0.36$\,M_\odot$ WD companion 
   instead  of  the 0.32$\,M_\odot$  WD.   This  could  be   because  the
   1.4G0.32C0.36D  case deposited  more kinetic  energy in  the ejecta
   than the  other two cases. The  final binary in this  case also has
   less remaining bound mass than in the other two cases (see Table
   \ref{tab:DWD}).
\end{itemize}

We  could  not  identify  any   other  initial  condition  that  could
discriminate why  the final  orbital separation  in the  two discussed
cases did  not follow the trend.   During a spiral-in, we  find that in
those two cases  the velocity at which the companion  plunges into the
envelope was higher  than in other cases, which is  consistent with the
ejecta taking  away more angular  momentum. However, what  causes this
deviation in the ejecta's angular momentum is not fully clear.

\begin{figure}
 \includegraphics[scale=0.4]{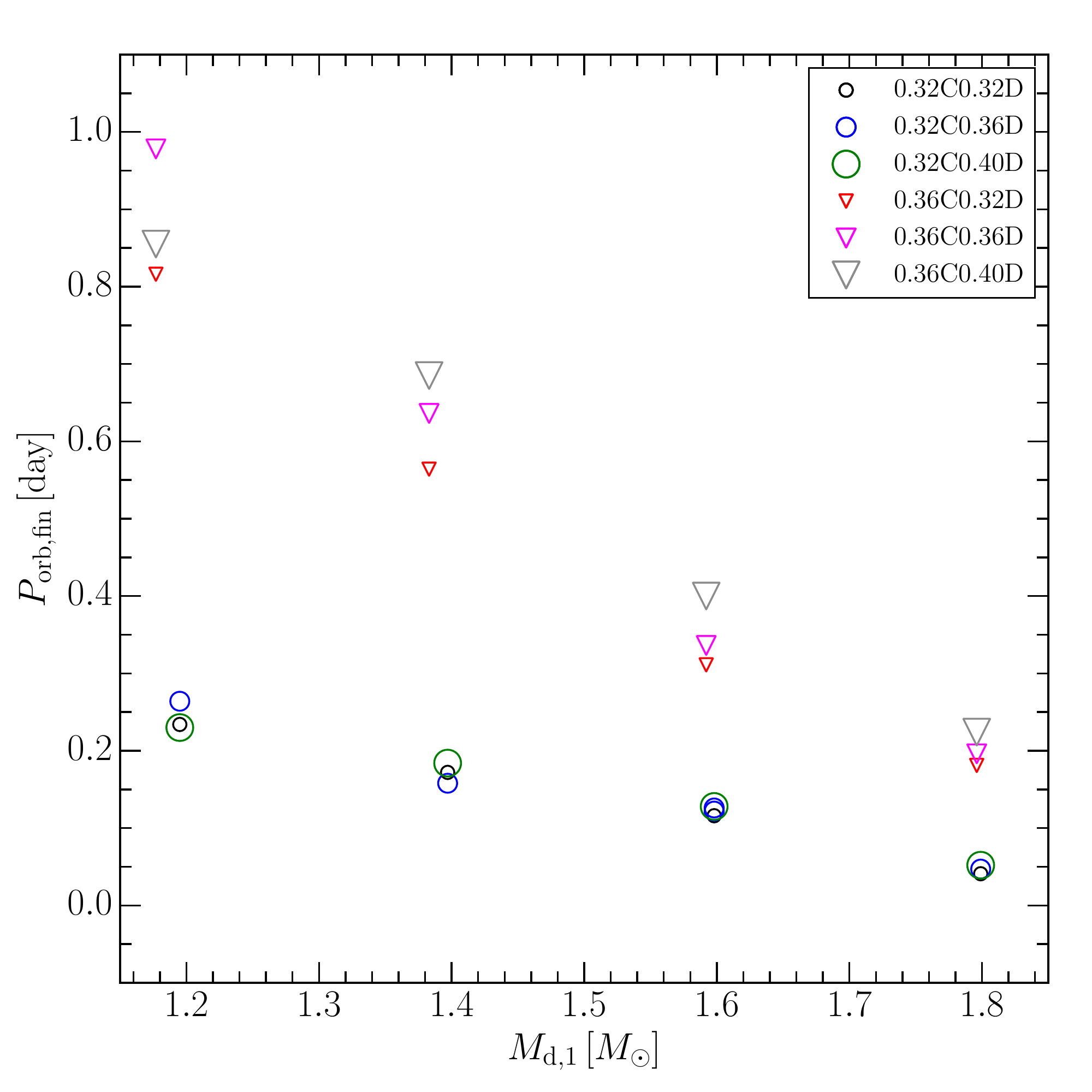}
 \caption{The final orbital periods, plotted against the initial RG mass.  The open circles correspond to simulations with $M_{\rm core,1}=0.32M_\odot$, while the open triangles correspond to $M_{\rm core,1}=0.36M_\odot$. The small, medium and big symbols are for companions with $0.32M_\odot$, $0.36M_\odot$, and $0.40M_\odot$, respectively.}
 \label{fig:fit_period_mass}
\end{figure}

Figure \ref{fig:fit_period_mass}  shows the final orbital  periods for
all the simulations, as a function of  the initial RG mass.  It can be
seen that qualitatively  there are two populations,  mainly defined by
the  mass of  the RG  core,  and with a smaller dependence  on the  mass of  the
companion.  In each of these two populations, the final orbital period
appears to depend almost linearly on the initial RG mass.

\begin{figure}
 \includegraphics[scale=0.4]{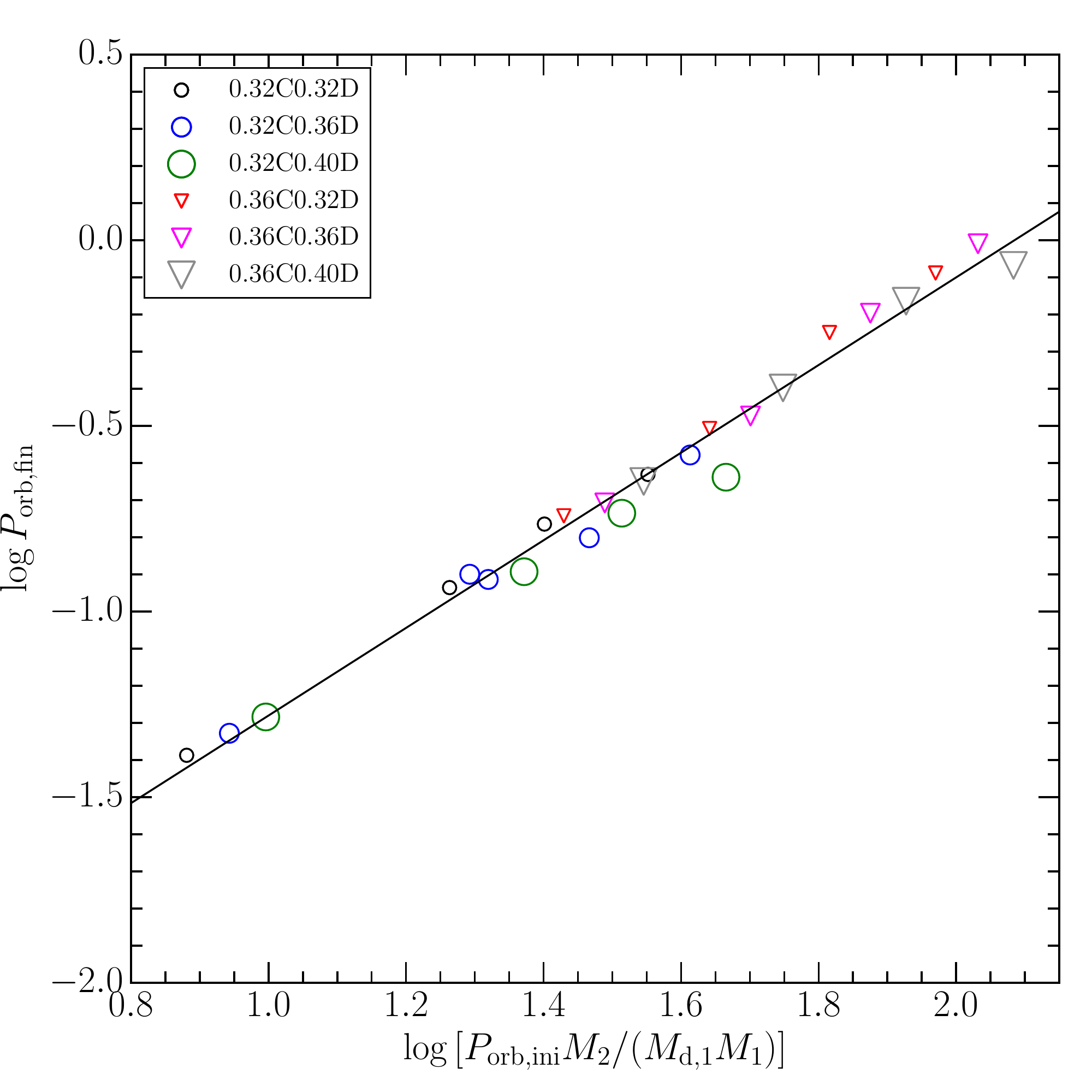}
 \caption{Final orbital periods. This plot contains all non-synchronized  simulations. The open circles correspond to simulations with $M_{1}\approx0.32M_\odot$, the open triangles correspond to $M_{1}\approx0.36M_\odot$. The small, medium and big symbols are for companions with $0.32M_\odot$, $0.36M_\odot$, and $0.40M_\odot$, respectively. The black solid line corresponds to the best fit for all the simulations (see Eq. \ref{eq:fittingline}). {The orbital periods are in days and the masses are in $M_\odot$.}}
 \label{fig:fit_period_period}
\end{figure}

Figure \ref{fig:fit_period_period} shows the  final orbital periods as
a function  of the  initial orbital period, the  initial RG
mass, the mass of the RG core  and the mass of  the companion. 
This appeared to produce the  relationship that can be expressed
as follows:

\begin{equation}
 P_{\rm orb,fin}=10^{-2.46\pm0.05}\left(P_{\rm orb,ini} \times \frac{M_2} {M_{\rm d,1} M_{1}}\right)^{1.18\pm0.04},
 \label{eq:fittingline}
 \end{equation}
\noindent Here $\pm$ indicates the standard error for each coefficient. The units for the quantities 
are  $M_\odot$ for all the masses, and days for periods.

\subsection{$\alpha_{\rm bind}\lambda$ formalism}

\begin{figure}
 \includegraphics[scale=0.4]{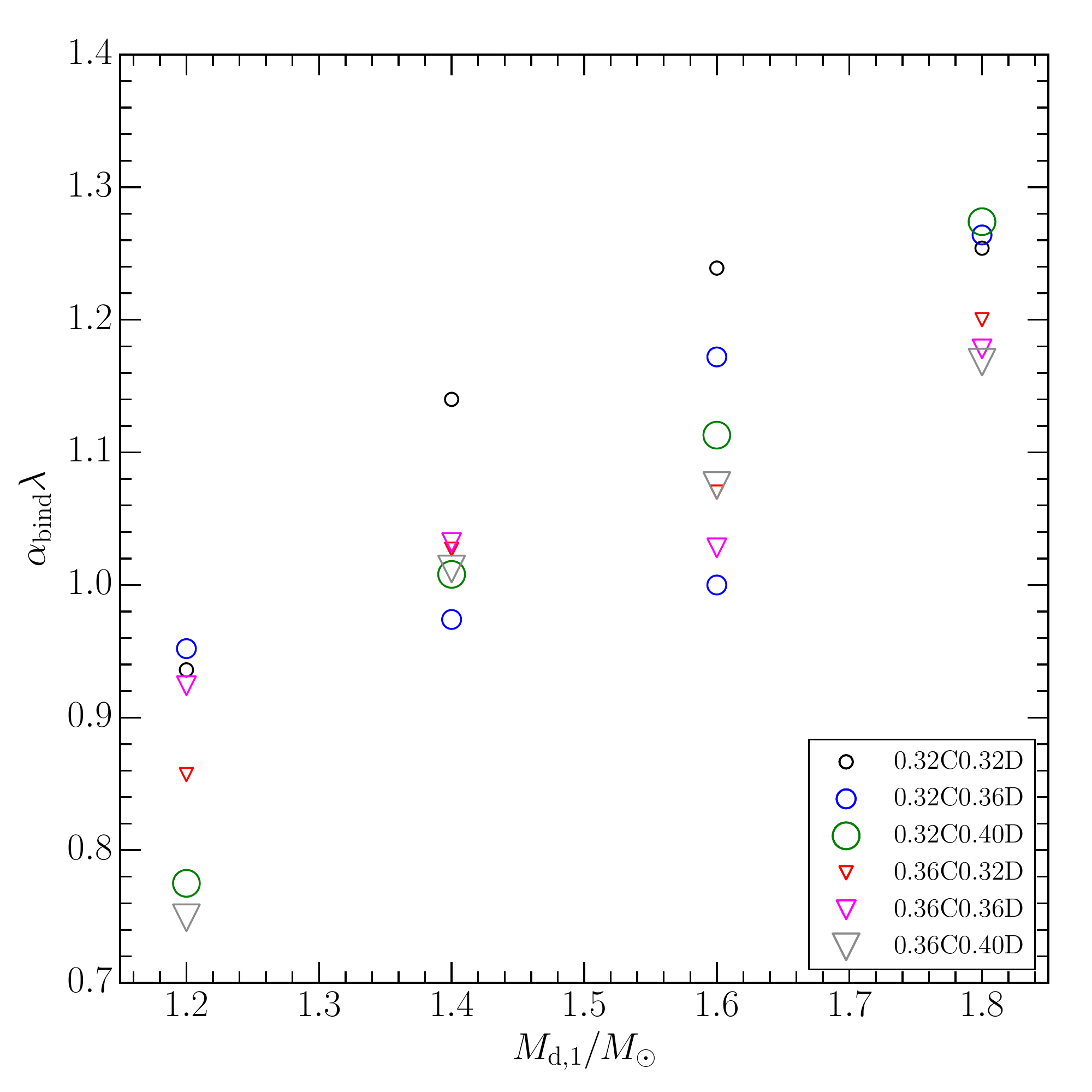}
 \caption{Values for $\alpha_{\rm bind}\lambda$ with respect to the initial RG mass.  The open circles correspond to simulations with $M_{\rm c,1}=0.32M_\odot$, while the open triangles correspond to $M_{\rm c,1}=0.36M_\odot$. The small, medium and big symbols are for companions with $0.32M_\odot$, $0.36M_\odot$, and $0.40M_\odot$, respectively. }
 \label{fig:alpha_bind_lam}
\end{figure}

In population  synthesis models, a crucial parameter is $\alpha_{\rm
  bind}\lambda$, which can be found from the results of our simulations as follows (see also Equation~\ref{eq:alphalambda}):
\begin{equation*}
 \alpha_{\rm bind}\lambda=-\frac{GM_{\rm d,1}(M_{\rm d,1}-M_{\rm c,1})}{R_{\rm rlof}\Delta E_{\rm orb}} \ .
\end{equation*}
Note that this quantity does not imply a separate consideration of the recombination energy as it is simply a fit to the standard energy formalism. Figure \ref{fig:alpha_bind_lam} shows  the behavior of $\alpha_{\rm bind}\lambda$ in our models.
Our best fit for $\alpha_{\rm  bind}\lambda$ with  the assumed multi-linear  regression model
is: 
\begin{equation}
 \alpha_{\rm  bind}\lambda=0.92+0.55\frac{M_{\rm
     d,1}}{M_\odot}-0.79\frac{M_{2}}{M_\odot}-1.19\frac{M_{\rm
     c,1}}{M_\odot}.
     \label{eq:alphalambdafit}
\end{equation}
This equation accurately represents all our models, and the maximum deviation
 between this equation and any data point of 0.13 (1.2G0.32C0.4D), 
and a minimum deviation of 0.002 (1.8G0.36C0.40D).

\subsection{Energy carried away by the ejecta } 

\begin{figure}
 \includegraphics[scale=0.4]{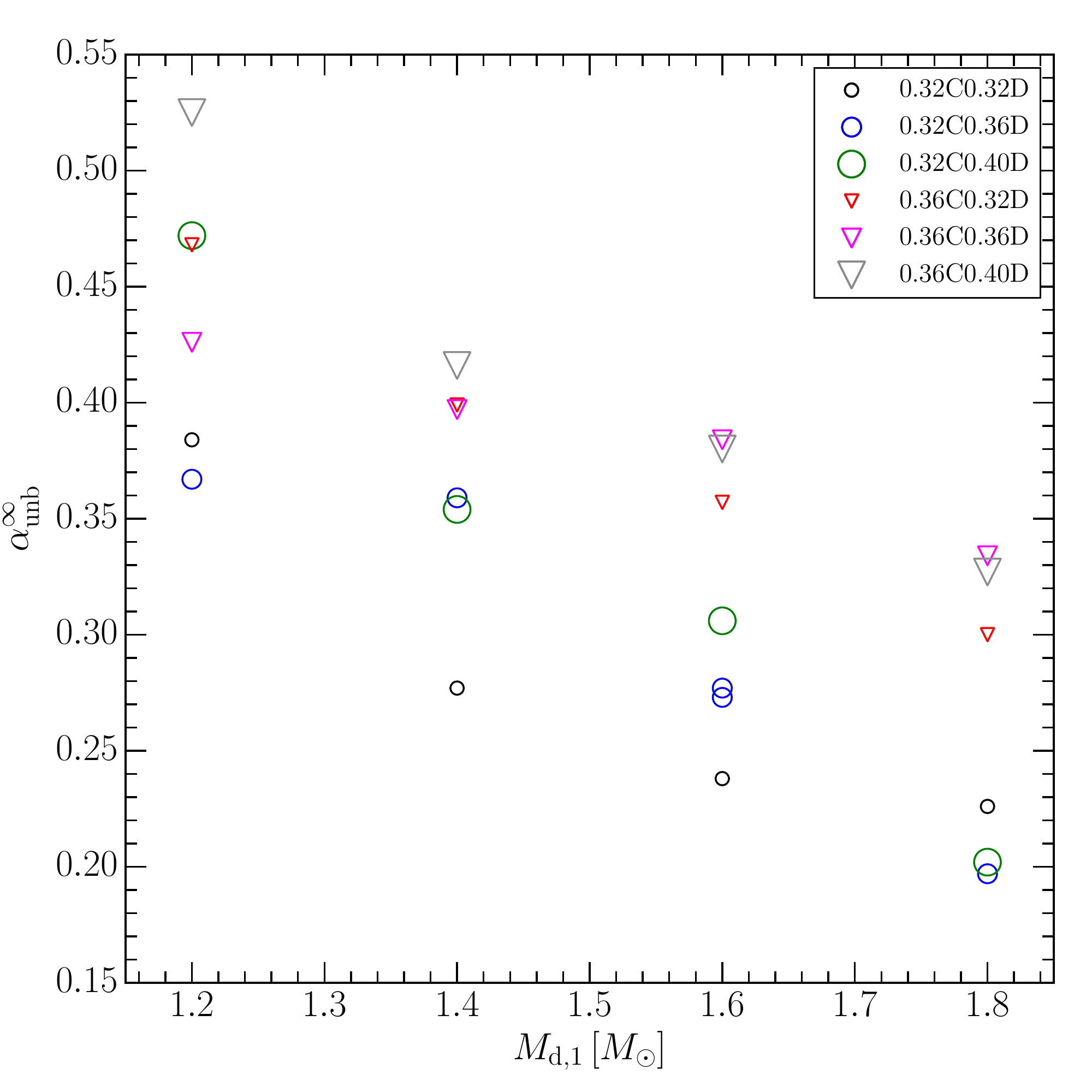}
 \caption{Values  for $\alpha_{\rm  unb}^\infty$ with  respect to  the
   initial RG mass.  The  open circles  correspond to
   simulations  with  $M_{\rm  core,1}=0.32M_\odot$,  while  the  open
   triangles  correspond to  $M_{\rm core,1}=0.36M_\odot$.  The small,
   medium  and  big symbols  are  for  companions with  $0.32M_\odot$,
   $0.36M_\odot$, and $0.40M_\odot$, respectively.}
 \label{fig:alpha_unb}
\end{figure}

The total  energy carried by the  ejecta is not negligible, and is comparable, within  an
order of magnitude, to  the initial binding energy of the
RG star.   Figure \ref{fig:alpha_unb} shows  the ratio of  the energy
taken away  by the  unbound material to  the released  orbital energy,
$\alpha_{\rm  unb}^\infty$.    It  can   be  seen   that  $\alpha_{\rm
  unb}^\infty$  decreases with  the  mass of  the  RG.  A  multilinear
regression that  uses all  the points  from the  simulations (assuming
that all the variables presented have linear trends with respect to each
other) gives the following dependence:

\begin{equation}
 \alpha_{\rm unb}^{\infty}=-\frac{E_{\rm tot,unb}^\infty}{\Delta E_{\rm orb}}=-0.16 - 0.30 \frac{M_{\rm d,1}}{M_\odot}+0.49\frac{M_{2}}{M_\odot}+2.27\frac{M_{\rm c,1}}{M_\odot}.
 %-0.244 \frac{M_{*,1}}{M_\odot}+0.706.
\end{equation}
We  note   that  this  equation   fits  all  our  models. The maximum deviation between this equation and any
point is found  to be 0.07 (1.2G0.32C0.40D), and the  minimum is 0.005
(1.4G0.36C0.32D).

\begin{figure}
 \includegraphics[scale=0.4]{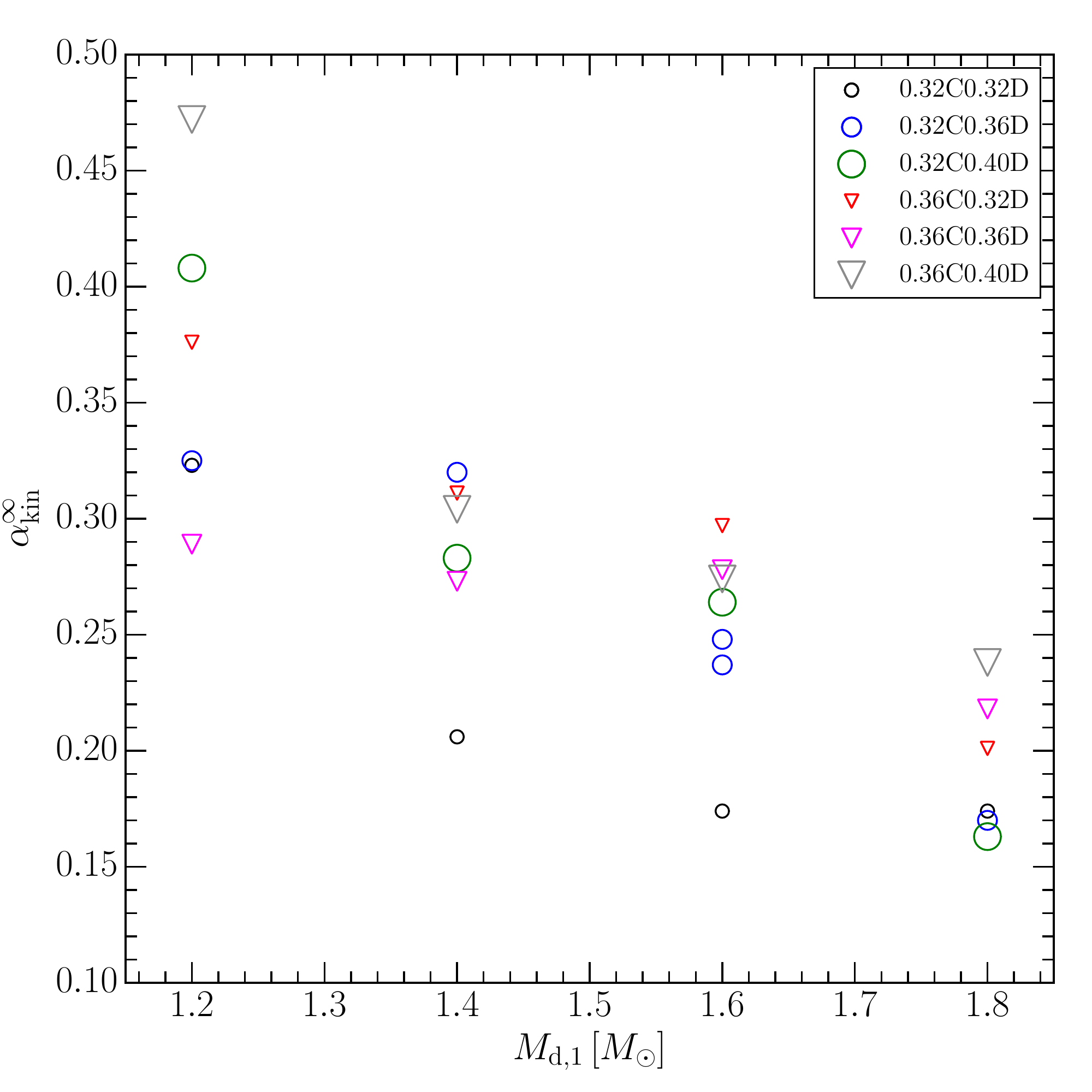}
 \caption{Values  for $\alpha_{\rm  kin}^\infty$ with  respect to  the
   initial RG mass.  The  open circles  correspond to
   simulations   with  $M_{\rm   c,1}=0.32M_\odot$,  while   the  open
   triangles  correspond  to  $M_{\rm  c,1}=0.36M_\odot$.  The  small,
   medium  and  big symbols  are  for  companions with  $0.32M_\odot$,
   $0.36M_\odot$, and $0.40M_\odot$, respectively.}
 \label{fig:alpha_kin}
\end{figure}

Table \ref{tab:DWDj}  shows $\alpha_{\rm kin}^{\infty}$,  which is
defined as the ratio of the kinetic energy taken away by the ejecta to the
released orbital energy.  Figure \ref{fig:alpha_kin} shows a monotonic
decrease of $\alpha_{\rm  kin}^{\infty}$ with the initial  mass of the
RG,  very similarly  to  $\alpha_{\rm  unb}^\infty$.  The  multilinear
fitting equation takes the following form
\begin{equation}
 \alpha_{\rm    kin}^{\infty}=-\frac{E_{\rm    kin,unb}^\infty}{\Delta
   E_{\rm     orb}}=0.20-0.26     \frac{M_{\rm     d,1}}{M_\odot}+0.44
 \frac{M_{2}}{M_\odot}+0.92 \frac{M_{\rm c,1}}{M_\odot},
\end{equation}
where   this  equation   fits   all  the   points   presented   in
Table~\ref{tab:DWDj}. The  maximum deviation between this  equation and
any listed value  in Table~\ref{tab:DWDj}  is 0.07  (1.4G0.32C0.32D),
while the minimum deviation is 0.0002 (1.2G0.32C0.32D).

The  potential energy  of the ejecta,  compared  to the  initial
potential  energy, is  not really  significant,  as in  all cases
$\alpha_{\rm pot}\lesssim0.04$.  The thermal  energy the ejecta still
has at  infinity, as compared to  the initial thermal energy  plus the
recombination energy, is several times  larger, albeit also limited
to $\alpha_{\rm th}\lesssim0.16$.  The thermal energy of the ejecta is
comparable  to  the  kinetic  energy of  the  ejecta,  therefore,  the
internal energy still plays a role in supporting the ongoing expansion
of the material even after all the material is unbound.

\begin{figure}
 \includegraphics[scale=0.4]{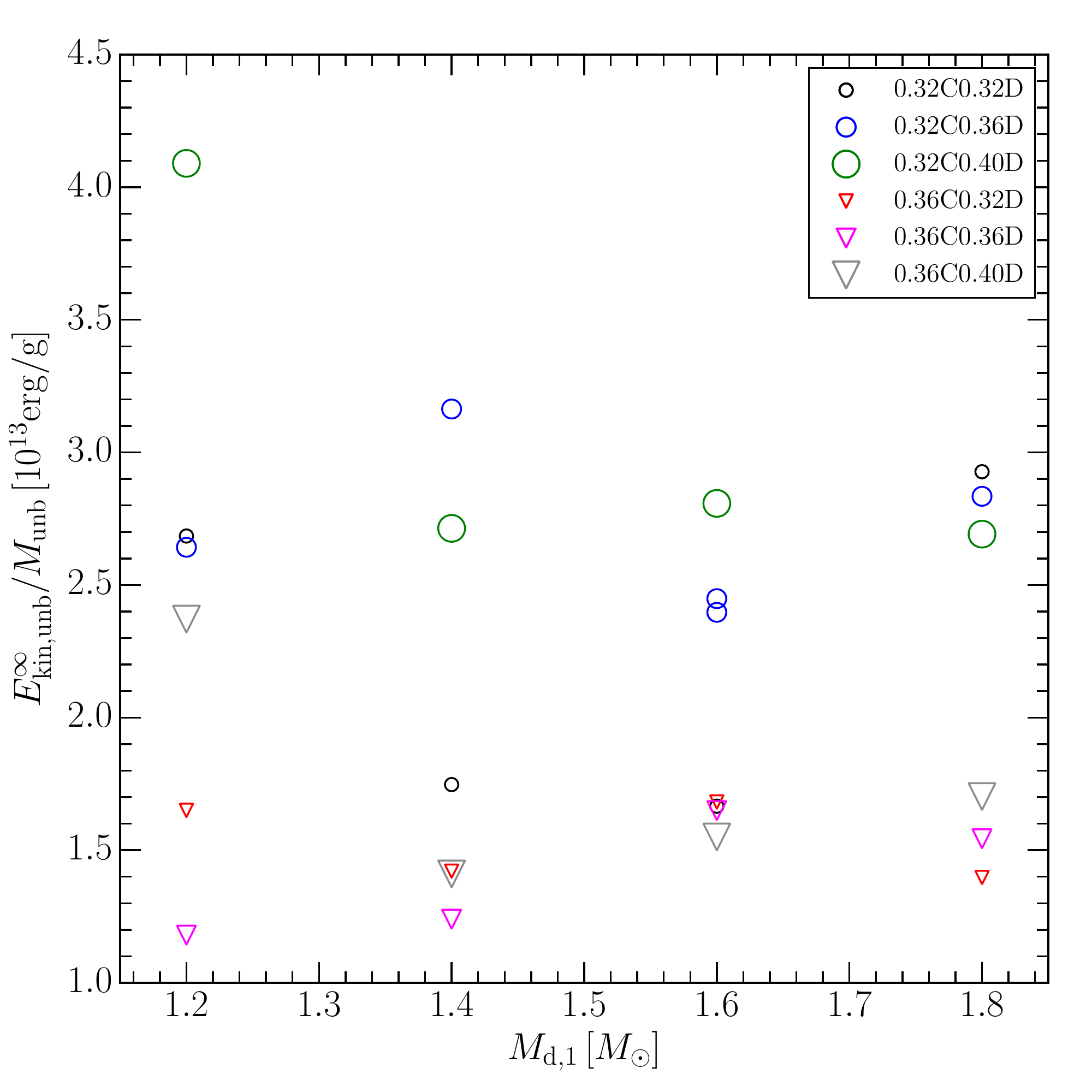}
 \caption{Values for $E_{\rm kin,unb}^\infty/M_{\rm unb}$ with respect
   to  the initial  RG mass.    The  open  circles
   correspond to simulations with $M_{\rm c,1}=0.32M_\odot$, while the
   open triangles correspond to  $M_{\rm c,1}=0.36M_\odot$. The small,
   medium  and  big symbols  are  for  companions with  $0.32M_\odot$,
   $0.36M_\odot$, and $0.40M_\odot$, respectively.}
 \label{fig:e_kin_inf}
\end{figure}

Figure \ref{fig:e_kin_inf}  shows how  the specific kinetic  energy of
the ejecta changes with  the initial mass of the RG.   We can see that
overall this energy  decreases as the RG mass increases,  but no clear
trend is observed.

\begin{figure}
 \includegraphics[scale=0.4]{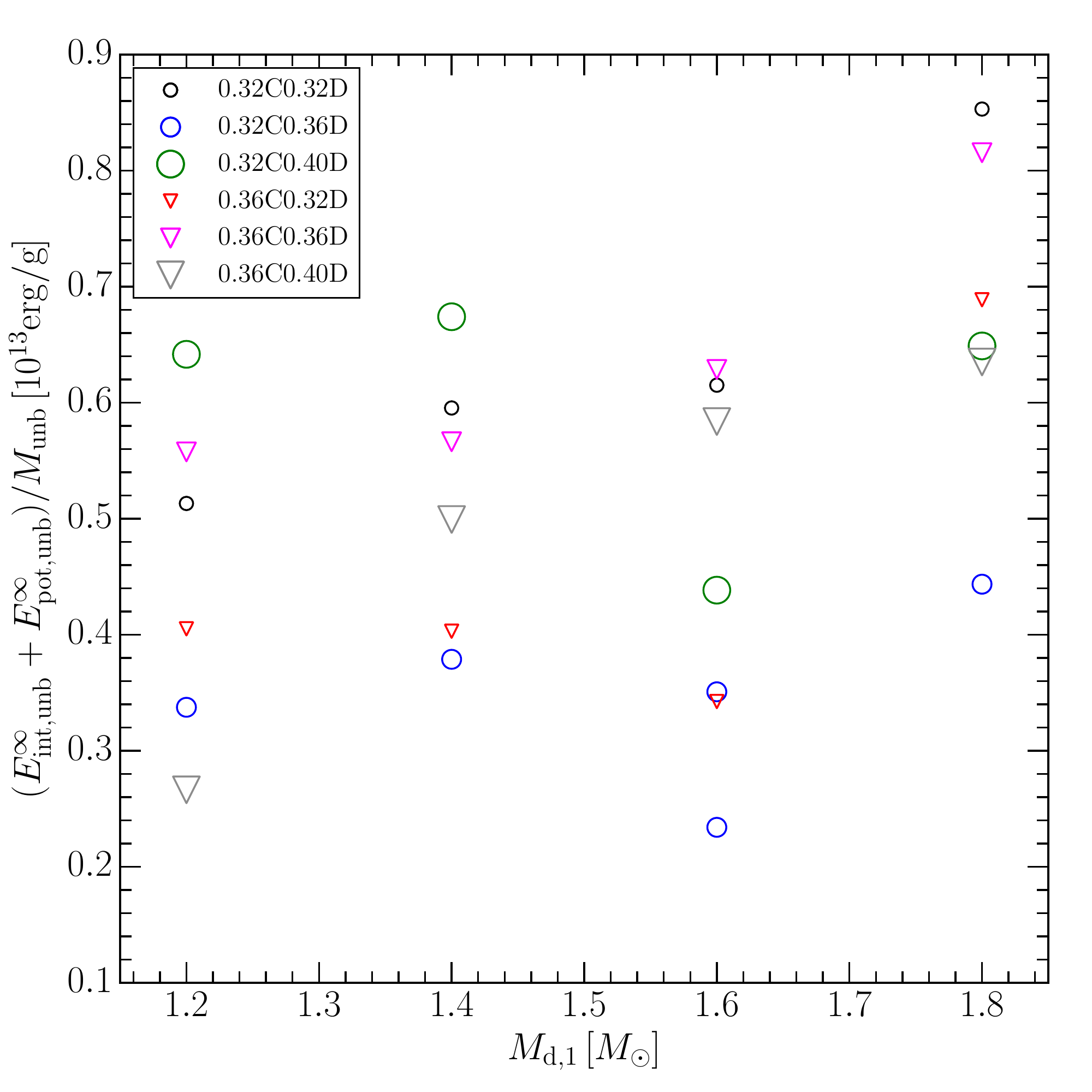}
 \caption{Values       for       $(E_{\rm       int,unb}^\infty+E_{\rm
     pot,unb}^\infty)/M_{\rm  unb}$ with  respect  to  the initial  RG
   mass.  The open circles correspond to simulations with
   $M_{\rm c,1}=0.32M_\odot$,  while the open triangles  correspond to
   $M_{\rm c,1}=0.36M_\odot$.  The small,  medium and big  symbols are
   for    companions    with   $0.32M_\odot$,    $0.36M_\odot$,    and
   $0.40M_\odot$, respectively. }
 \label{fig:e_pot_inf}
\end{figure}

Figure  \ref{fig:e_pot_inf} shows  how the  sum of  the potential  and
thermal specific energies  of the ejecta changes with the  mass of the
RG.   We cannot  really see  a  trend, except  that for  RGs with  the
initial mass of  $1.8\, M_\odot$ this quantity is higher  than for the
rest.  Note  that this  quantity is always  smaller than  the specific
kinetic energy of the ejecta.

\section{Discussion}
\label{sec:conclusion}

To study the formation of a DWD binary via a CEE, we have simulated 25
three-dimensional hydrodynamical  interactions between  a low-mass  RG 
and  a WD  companion.  We  considered for  the initial  masses of  the
low-mass RG star 1.2,  1.4, 1.6, or 1.8 $M_\odot$, with  a He core of
0.32 or 0.36 $M_\odot$, and WD companions with masses 0.32, 0.36, or
0.40  $M_\odot$.  We  find that  in  all the  cases, a  DWD binary  is
formed, most of the envelope  is ejected, and only a few SPH
particles remain  bound to the  binary in some cases (the bound  mass is
less  than 0.06\%  of the  initial  envelope mass).   The envelope  is
ejected on the dynamical time-scale of the expanded envelope.

Our results show that the standard energy formalism should be modified
to take into account: (i) the energy that is taken away by the ejecta,
as it  is a substantial fraction  of the released orbital  energy, and
(ii) the recombination energy, which  plays a crucial role in ejection
of the formed circumbinary envelope. The role of the recombination energy for
the CEE with a  low-mass RG donor is not that it  is necessary for the
overall energy budget, as none of the considered systems were expected
to  merge   by  the  standard   energy  formalism,  but   because  the
recombination occurs exactly  at the time when the  shrunk binary is
no longer capable of transfering  its orbital energy to the expanded
envelope.

For future population synthesis studies, we  provide three ways in which our
results can be used.

First   of    all,   we   provide   a    fitting   formula   (Equation
\ref{eq:fittingline})  that  relates  the final  orbital  period,  the
initial orbital  period, the companion  mass, the initial RG  mass and
the RG  core.  The RG  radius and its core  mass are coupled  for each
donor mass (these can be  found using single stellar evolution tracks).
The initial  RG mass, its  radius and  the initial orbital  period are
also related \citep[e.g., by using the Roche lobe radius approximation
  from][]{1983ApJ...268..368E}. Therefore our fitting formula provides
the  relation between  the observed  parameters --the  post-CE orbital
period and the observed  masses of  both WDs  -- and the  RG mass  and radius
before  the CE.   And  vice versa,  a population  study  could use  this
fitting formula to obtain the post-CE orbital period from known binary
parameters at the start of a CE.

\begin{figure}
 \includegraphics[scale=0.4]{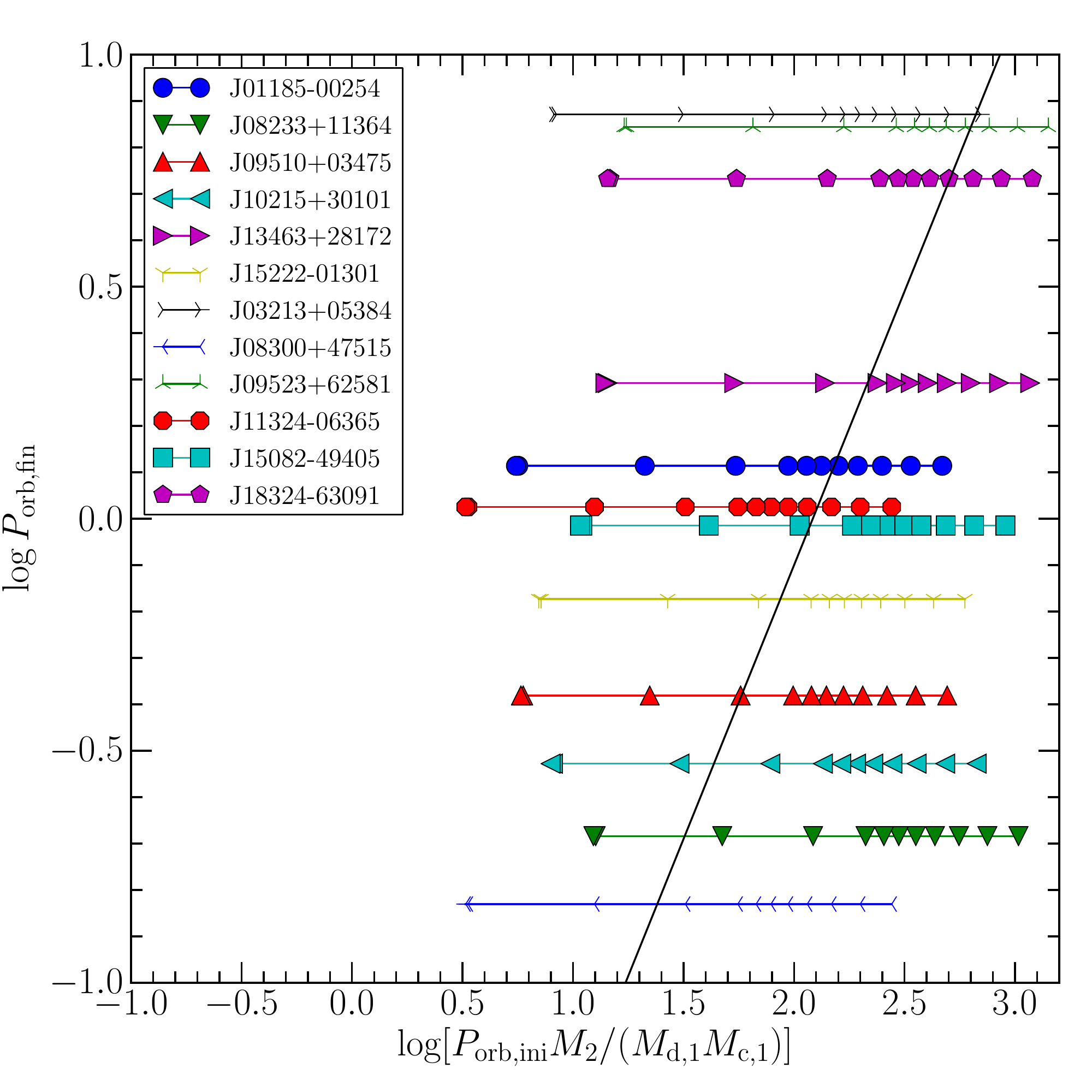}
 \caption{Final orbital periods of  sdB stars  \citep[from][]{2015A&A...576A..44K} and
   their pre-CE conditions,     as    inferred from
   Equation \ref{eq:fittingline}.  The solid black line is the relation
   provided  by  Equation \ref{eq:fittingline}.   The  crossing
   horizontal  lines show  the positions  for different  initial donor
   masses, using observed  older WD masses.  Solid symbols  are for RG
   donors (ZAMS masses from 1.0 to 1.8 $M_\odot$ with an increment
   of $0.1\,M_\odot$, with the lowest mass  on the right.) The RG donors
   are selected when their cores  are $0.47 M_\odot$.  Open symbols are
   for  AGB  donors (ZAMS  masses  1.8,  1.9  and 2.0  $M_\odot$,  the
   smallest on  the right). The AGB  donors are  selected when their cores are
   $0.53\,M_\odot$).  }
 \label{fig:sdb_binaries}
\end{figure}

\begin{table*}
\begin{minipage}{175mm}
 \caption{sdB binary predictions}
 \label{tab:sdBbin}
 \begin{center}
 \begin{tabular}{lcccllcll}
  \hline
  Object &$M_{\rm ZAMS}$ &$M_{\rm d,1}$ & $M_{\rm c,1}$ &$R_{\rm d,1}$& $M_{2}$ &$P_{\rm orb,ini}$ & $P_{\rm orb,fin}$&Stage \\
  \hline
  J08300+47515&1.80&1.75&0.53&110&0.14&212.03&0.15&AGB\\
  J08233+11364&1.90&1.87&0.53&64 &0.44&106.94&0.21&AGB\\
  J10215+30101&1.90&1.87&0.53&64 &0.30&100.42&0.30&AGB\\
  J09510+03475&1.80&1.75&0.53&110&0.23&229.56&0.43&AGB\\
  J15222-01301&1.80&1.75&0.53&110&0.27&235.69&0.67&AGB\\
  J15082-49405&1.80&1.75&0.53&110&0.39&250.38&0.97&AGB\\
  J11324-06365&1.30&1.20&0.47&157&0.14&461.31&1.06&RG\\
  J01185-00254&1.40&1.32&0.47&153&0.22&449.81&1.30&RG\\
  J13463+28172&1.70&1.67&0.47&139&0.49&378.87&1.96&RG\\
  J18324-63091&1.30&1.20&0.47&157&0.50&565.64&5.40&RG\\
  J09523+62581&1.30&1.20&0.47&157&0.58&577.70&6.98&RG\\
  J03213+05384&1.00&0.78&0.47&176&0.31&824.86&7.43&RG\\
  \hline
 \end{tabular}
 \end{center}
 \medskip
 $M_{\rm ZAMS}$, $M_{\rm d,1}$, $M_{\rm c,1}$, and $M_2$ are the ZAMS mass of the donor from our parameter space, inferred initial donor mass right before the CEE, 
 mass of the donor's core  (which is assumed to become a sdB star after the CEE), and the companion mass \citep[assumed to be the minimum companion mass from ][]{2015A&A...576A..44K}, respectively. All the masses are in $M_{\odot}$. $R_{\rm d,1}$ is the inferred radius of the donor star right 
 before the CEE  in $R_{\odot}$. $P_{\rm orb,ini}$, and $P_{\rm orb,ini}$ are the inferred initial orbital period, and the observed orbital period, respectfully, in 
 day. Stage is the evolutionary stage of the donor right before the CEE, AGB star or RG star.
 \end{minipage}
\end{table*}

Let us consider how this fitting  formula can be used to interpret 
observed  post-CE  binaries.   For  example,  take the  WD
1101+364,  which has  observed  parameters best matching  the set  of
models  we  have  calculated:  $P_{\rm  orb,fin}=0.145$  day,  $M_{\rm
  1}=0.31\,     M_\odot$     and    $M_{\rm     2}=0.36\,     M_\odot$
\citep{1995MNRAS.275L...1M}.  To find the radius of the donor at the time
when it had  a core of the same  mass as a younger WD  in the observed
sample,    we   used    parameterized    evolutionary   tracks    from
\citet{2000MNRAS.315..543H}.   For  WD~1101+364, the  fitting  formula
predicts  a pre-CE  donor mass  of $1.5~M_\odot$,  and pre-CE  orbital
period of 33  days. We note that more detailed  studies devoted to the
simulations  specifically of  WD~1101+364  gave a similar pre-CE  donor
mass, $1.5~M_\odot$ \citet{nandez2015}.  We note that since the pre-CE
radius is a strong function of  the core mass, uncertainty in the mass
of  a  younger  WD  leads  to   a  large  uncertainty  on  the  pre-CE
donor. E.g.,  if the  mass is  only a  bit smaller,  $M_{\rm 1}=0.29\,
M_\odot$, and  the companion's mass is $M_{\rm 2}=0.33\, M_\odot$ (defined by the  observed mass
ratio of 0.87), then the donor would rather
have an initial mass of $1.3M_\odot$ and a pre-CE orbital period of
26 days.

The second observable  type of post-CE binaries for which  we can test
the fitting equation are hot subdwarf B stars (sdBs).  These stars are
hot core  helium-burning stars with  masses around $0.5  M_\odot$.  We
note that the post-CE remnants of this  mass are beyond the set of our
current simulations (modeling a CEE with a more evolved donor requires
substantially  more  GPU  time  than is  available  at  existing
Compute/Calcul Canada facilities, and  therefore is not feasible yet),
but we will try to look at the post-CE binaries  to see
if we can place any constraint on  their past.  We use  12 sdB binaries
for which \citet{2015A&A...576A..44K} have  found orbital periods, and
inferred  the  minimum companion  masses  in  these systems  from  the
assumption of a  canonical mass of $0.47\,M_\odot$ for  the sdB stars.
In  Figure~\ref{fig:sdb_binaries}  we  show  the  predictions  of our 
fitting formula.  In addition to checking RG donors, we also took
into  account  AGB   donors.  A prediction for a post-CE  outcome for an
AGB donor  can not be  fully trusted, as  the fitting formula  may not
work well  for them; both $\lambda$  in the donor's envelope,  and the
fraction of  the recombination energy  in the total binding  energy of
the initial envelope, are not the same as in the case of a low-mass RG.
However, it  is important that at  least half of  the considered sdB
binary systems can be better explained by a RG donor.

Table~\ref{tab:sdBbin} summarizes  the  possible progenitors  for
each sdB binary. We note that we list the values for the closest point
on  the  Figure~\ref{fig:sdb_binaries},   not  the  exact  intersection
between the model and the line with constant final orbital period; the
donor mass for each case will not change much from the listed value in
the  Table, only  the  radius of  the donor  and  its initial  orbital
period will change.   We   can  see  from  Table~\ref{tab:sdBbin} or Figure~\ref{fig:sdb_binaries} that  the
evolutionary  stage of  the donor  star can  be associated  with the
final orbital  period. For  $P_{\rm orb,fin}\lesssim1$ day,  the donor
star is likely a (relatively more massive) AGB star, while for $P_{\rm
  orb,fin}\gtrsim1$  day,  the donor  star  is  more  likely to  be  a
(relatively less massive) RG.

For  a  second way  to  use  our results  --  as  in the  population
synthesis  studies that  use  $\alpha_{\rm bind}\lambda$-formalism  to
find  the outcome  of  a CEE  -- we  supply  the parameterization  that
directly provides  the $\alpha_{\rm bind}\lambda$ value, as  a function of
the initial donor mass, its core mass and the companion mass.  We
note that in no case  $\alpha_{\rm bind}\lambda>1.3$, and our
maximum $\alpha_{\rm bind}<1.03$ (we remind that $\alpha_{\rm bind}$ more than 1 implies that an energy additional to the orbital
energy was used, in our case it is the recombination energy).  Some past population synthesis
studies  have   considered  $\alpha_{\rm   bind}\lambda=2$  \citep[see
  e.g.][]{2012A&A...546A..70T} for  all CEEs  leading to  DWD binary 
formation, but  the results  of our  simulations do not confirm 
that such  a very high  value is plausible, at  least in the case  of CEEs
with low-mass RG donors.
 
And, finally, we give the preferred way  to use our results,  which is the
most trusted method  when one wants to extrapolate  our result outside
of  the parameter  space we considered.  We  advise  population
synthesis  studies  to  use  the  energy  conservation  equation  that
accounts  for all energy  sinks  and  sources. 
In the energy conservation equation,  all  initially
available recombination energy can be  used as an energy source. (Note
that this  statement is not  yet fully justified to  extend our results for low-mass
giants to the  case of more massive or more  evolved donors, and shall
require further studies.) 
 The ejected  material can take away 
20\% to 40\%  of the released orbital energy, both  as thermal energy
and as  kinetic energy,  and this  is an energy  loss.  It  is this
energy  that powers those Luminous  Red Novae  which are
produced by a CEE \citep{2013Sci...339..433I}.  For these energy losses
we provided  a fitting formula.   Then the  CEE outcomes can  be found
using the {\bf revised energy formalism} as follows:

\begin{equation}
(E_{\rm orb,ini}-E_{\rm orb,fin})  (1-\alpha_{\rm unb}^{\infty}) +  E_{\rm bind}  
+ \eta (M_{\rm d,1}- M_{\rm c,1})  =0 \ .
\end{equation}

\section*{Acknowledgments}

JLAN acknowledges CONACyT for its support. 
NI thanks NSERC Discovery and Canada Research Chairs Program. 
The authors thank Craig Heinke for checking the English in the manuscript.
This research has been enabled by the use of computing resources provided 
by WestGrid and Compute/Calcul Canada.

\bibliographystyle{mn2e}
\bibliography{references}

\begin{thebibliography}{}

\bibitem[\protect\citeauthoryear{{de Kool}}{{de
  Kool}}{1990}]{1990ApJ...358..189D}
{de Kool} M.,  1990, \apj, 358, 189

\bibitem[\protect\citeauthoryear{{Eggleton}}{{Eggleton}}{1971}]{1971MNRAS.151..351E}
{Eggleton} P.~P.,  1971, \mnras, 151, 351

\bibitem[\protect\citeauthoryear{{Eggleton}}{{Eggleton}}{1972}]{1972MNRAS.156..361E}
{Eggleton} P.~P.,  1972, \mnras, 156, 361

\bibitem[\protect\citeauthoryear{{Eggleton}}{{Eggleton}}{1983}]{1983ApJ...268..368E}
{Eggleton} P.~P.,  1983, \apj, 268, 368

\bibitem[\protect\citeauthoryear{{Gaburov}, {Lombardi} Jr. \& {Portegies
  Zwart}}{{Gaburov} et~al.}{2010}]{2010MNRAS.402..105G}
{Gaburov} E.,  {Lombardi} Jr. J.~C.,    {Portegies Zwart} S.,  2010, \mnras,
  402, 105

\bibitem[\protect\citeauthoryear{{Glebbeek}, {Pols} \& {Hurley}}{{Glebbeek}
  et~al.}{2008}]{2008A&A...488.1007G}
{Glebbeek} E.,  {Pols} O.~R.,    {Hurley} J.~R.,  2008, \aap, 488, 1007

\bibitem[\protect\citeauthoryear{{Han}, {Podsiadlowski}, {Maxted}, {Marsh} \&
  {Ivanova}}{{Han} et~al.}{2002}]{2002MNRAS.336..449H}
{Han} Z.,  {Podsiadlowski} P.,  {Maxted} P.~F.~L.,  {Marsh} T.~R.,    {Ivanova}
  N.,  2002, \mnras, 336, 449

\bibitem[\protect\citeauthoryear{{Hernquist} \& {Katz}}{{Hernquist} \&
  {Katz}}{1989}]{1989ApJS...70..419H}
{Hernquist} L.,  {Katz} N.,  1989, \apjs, 70, 419

\bibitem[\protect\citeauthoryear{{Hurley}, {Pols} \& {Tout}}{{Hurley}
  et~al.}{2000}]{2000MNRAS.315..543H}
{Hurley} J.~R.,  {Pols} O.~R.,    {Tout} C.~A.,  2000, \mnras, 315, 543

\bibitem[\protect\citeauthoryear{{Iben} Jr. \& {Tutukov}}{{Iben} \&
  {Tutukov}}{1984}]{1984ApJS...54..335I}
{Iben} Jr. I.,  {Tutukov} A.~V.,  1984, \apjs, 54, 335

\bibitem[\protect\citeauthoryear{{Ivanova}, {Justham}, {Avendano Nandez} \&
  {Lombardi}}{{Ivanova} et~al.}{2013}]{2013Sci...339..433I}
{Ivanova} N.,  {Justham} S.,  {Avendano Nandez} J.~L.,    {Lombardi} J.~C.,
  2013, Science, 339, 433

\bibitem[\protect\citeauthoryear{{Ivanova}, {Justham}, {Chen}, {De Marco},
  {Fryer}, {Gaburov}, {Ge}, {Glebbeek}, {Han}, {Li}, {Lu}, {Marsh},
  {Podsiadlowski}, {Potter}, {Soker}, {Taam}, {Tauris}, {van den Heuvel} \&
  {Webbink}}{{Ivanova} et~al.}{2013}]{2013A&ARv..21...59I}
{Ivanova} N.,  {Justham} S.,  {Chen} X.,  {De Marco} O.,  {Fryer} C.~L.,
  {Gaburov} E.,  {Ge} H.,  {Glebbeek} E.,  {Han} Z.,  {Li} X.-D.,  {Lu} G.,
  {Marsh} T.,  {Podsiadlowski} P.,  {Potter} A.,  {Soker} N.,  {Taam} R.,
  {Tauris} T.~M.,  {van den Heuvel} E.~P.~J.,    {Webbink} R.~F.,  2013, \aapr,
  21, 59

\bibitem[\protect\citeauthoryear{{Ivanova}, {Justham} \&
  {Podsiadlowski}}{{Ivanova} et~al.}{2015}]{2015MNRAS.447.2181I}
{Ivanova} N.,  {Justham} S.,    {Podsiadlowski} P.,  2015, \mnras, 447, 2181

\bibitem[\protect\citeauthoryear{{Ivanova} \& {Nandez}}{{Ivanova} \&
  {Nandez}}{2016}]{ivanova16}
{Ivanova} N.,  {Nandez} J.~L.,  2016, Submitted to MNRAS, \, \

\bibitem[\protect\citeauthoryear{{Kupfer}, {Geier}, {Heber}, {{\O}stensen},
  {Barlow}, {Maxted}, {Heuser}, {Schaffenroth} \& {G{\"a}nsicke}}{{Kupfer}
  et~al.}{2015}]{2015A&A...576A..44K}
{Kupfer} T.,  {Geier} S.,  {Heber} U.,  {{\O}stensen} R.~H.,  {Barlow} B.~N.,
  {Maxted} P.~F.~L.,  {Heuser} C.,  {Schaffenroth} V.,    {G{\"a}nsicke} B.~T.,
   2015, \aap, 576, A44

\bibitem[\protect\citeauthoryear{{Livio} \& {Soker}}{{Livio} \&
  {Soker}}{1988}]{1988ApJ...329..764L}
{Livio} M.,  {Soker} N.,  1988, \apj, 329, 764

\bibitem[\protect\citeauthoryear{{Lombardi} Jr., {Holtzman}, {Dooley},
  {Gearity}, {Kalogera} \& {Rasio}}{{Lombardi}
  et~al.}{2011}]{2011ApJ...737...49L}
{Lombardi} Jr. J.~C.,  {Holtzman} W.,  {Dooley} K.~L.,  {Gearity} K.,
  {Kalogera} V.,    {Rasio} F.~A.,  2011, \apj, 737, 49

\bibitem[\protect\citeauthoryear{{Marsh}}{{Marsh}}{1995}]{1995MNRAS.275L...1M}
{Marsh} T.~R.,  1995, \mnras, 275, L1

\bibitem[\protect\citeauthoryear{Nandez, Ivanova \& J.~C.~Lombardi}{Nandez
  et~al.}{2014}]{0004-637X-786-1-39}
Nandez J. L.~A.,  Ivanova N.,    J.~C.~Lombardi J.,  2014, The Astrophysical
  Journal, 786, 39

\bibitem[\protect\citeauthoryear{{Nandez}, {Ivanova} \& {Lombardi}}{{Nandez}
  et~al.}{2015}]{nandez2015}
{Nandez} J.~L.~A.,  {Ivanova} N.,    {Lombardi} J.~C.,  2015, \mnras, 450, L39

\bibitem[\protect\citeauthoryear{{Paxton}, {Bildsten}, {Dotter}, {Herwig},
  {Lesaffre} \& {Timmes}}{{Paxton} et~al.}{2011}]{2011ApJS..192....3P}
{Paxton} B.,  {Bildsten} L.,  {Dotter} A.,  {Herwig} F.,  {Lesaffre} P.,
  {Timmes} F.,  2011, \apjs, 192, 3

\bibitem[\protect\citeauthoryear{{Portegies Zwart} \& {Meinen}}{{Portegies
  Zwart} \& {Meinen}}{1993}]{1993A&A...280..174P}
{Portegies Zwart} S.~F.,  {Meinen} A.~T.,  1993, \aap, 280, 174

\bibitem[\protect\citeauthoryear{{Toonen}, {Nelemans} \& {Portegies
  Zwart}}{{Toonen} et~al.}{2012}]{2012A&A...546A..70T}
{Toonen} S.,  {Nelemans} G.,    {Portegies Zwart} S.,  2012, \aap, 546, A70

\bibitem[\protect\citeauthoryear{{Tutukov} \& {Yungelson}}{{Tutukov} \&
  {Yungelson}}{1981}]{1981NInfo..49....3T}
{Tutukov} A.~V.,  {Yungelson} L.~R.,  1981, Nauchnye Informatsii, 49, 3

\bibitem[\protect\citeauthoryear{{Tutukov} \& {Yungelson}}{{Tutukov} \&
  {Yungelson}}{1988}]{1988SvAL...14..265T}
{Tutukov} A.~V.,  {Yungelson} L.~R.,  1988, Soviet Astronomy Letters, 14, 265

\bibitem[\protect\citeauthoryear{{van den Heuvel}}{{van den
  Heuvel}}{1976}]{1976IAUS...73...35V}
{van den Heuvel} E.~P.~J.,  1976, in {Eggleton} P.,  {Mitton} S.,   {Whelan}
  J.,  eds, Structure and Evolution of Close Binary Systems Vol.~73 of IAU
  Symposium, {Late Stages of Close Binary Systems}.
p.~35

\bibitem[\protect\citeauthoryear{{Webbink}}{{Webbink}}{1984}]{1984ApJ...277..355W}
{Webbink} R.~F.,  1984, \apj, 277, 355

\end{thebibliography}

\label{lastpage}

\end{document}